\documentclass[11pt]{aa}
\usepackage[english]{babel}
\usepackage{supertab}
\usepackage{graphicx}

\newcommand{\alet}{Astron. Let. }
\newcommand{\mnras}{Mon. Not. R. Astron. Soc. }

\newcommand{\aaa}{Astron and Astrophys.}
\newcommand{\aas}{Astron and Astrophys. Suppl.}
\newcommand{\bsao}{Bull. Spec. Astrophys. Observ }

\onecolumn
\begin{document}

\markboth{Chentsov et al.}{Spectral atlas of MWC\,314 and IRC\,+10420}

\title{Spectral atlas of two peculiar supergiants: MWC\,314 and IRC\,+10420}
\author{E.L. Chentsov \and V.G. Klochkova \and N.S. Tavolganskaya}
\institute{Special Astrophysical Observatory RAS, Nizhnij Arkhyz,  369167 Russia}

\date{\today}

\abstract 
  {An  atlas  of  spectra  of  the  unique high luminosity objects
  MWC\,314   and   IRC\,+10420   taken   with   the  CCD  echelle
  spectrograph  at  the  prime focus of the 6\,m telescope of SAO
  is  presented. About 320 emission and absorption features of stellar,
  circumstellar  and  interstellar  nature  are identified in the
  interval   between   4790   and  7520\,\AA\AA.  Their  relative
  intensities   and   radial   velocities   are   given  and  the
  distinctions of their profiles are discussed.}
  
\authorrunning{\it Chentsov et al.}
\titlerunning{\it Spectral atlas of MWC\,314 and IRC\,+10420 }
 
\maketitle

\section{Introduction}

  The  purpose  of  the  paper  is  two-fold: to represent in the
  graphical  and  tabulated  forms the spectra which may prove to
  be  helpful  in  the  studies of various Be and Ae stars and to
  isolate  the  characteristic  properties  that have to be taken
  into     account     in    constructing    their    photosphere
  (pseudophotosphere) and envelope models.

The spectra of MWC\,314 and IRC\,+10420 are brought together for the
following reasons.
\begin{itemize}
\item In spite of the fact that the sets of photospheric absorption lines in
the objects and, accordingly, their spectral classes (given in Table\,1) are
 different, their emission spectra bear resemblance to one another. In both
cases one and the same permitted and forbidden components of iron group ion
lines dominate. The profiles of these lines are, however, different. When
going from MWC\,314 to IRC\,+10420, some emission lines get narrower, others
acquire absorption components or even transform fully to absorptions.
This helps in identification, especially in clearing up the composition of
blends.
\item The nature and evolutionary status of the stars are not quite clear.
Both of them are super- or even hypergiants, they are surrounded with envelopes
of complex structure and kinematics (Miroshnichenko et al., 1998, Klochkova
et al., 1997, Oudmaijer, 1998).
\item They are of value also for refining the structure of the Galaxy since
they are close to each other on the sky, in the north-western part of Aquila
(Table\,1). This Milky Way region is deficient in bright O stars and supergiants
since it corresponds to a direction between the local arm and the Car--Sgr arm.
\end{itemize}

\begin{table*}[hbtp]
\begin{center}
\caption{Main data of the objects}
\medskip
\begin{tabular}{clcccc}
\hline
  Name     &$\alpha, \delta$ (2000)& l, b                & V          & Sp & Date    \\
\hline
  MWC 314  & $\rm 19^h21^m34^s.1$  &$\rm 49^o.6, + 0^o.3$& $\rm 9^m.9$& B3 & 22.11.97\\
(V1429 Ag1)& $\rm 14^o52'56''$ &                     &            &    &         \\
           &                       &                     &            &    &         \\
IRC +10420 & $\rm 19^h26^m48^s.1$  &$\rm 47^o.1, - 2^o.5$&$\rm 11^m.0$& A5 & 19.05.97\\
(V1302 Ag1)& $\rm 11^o21'17''$ &                     &            &    &         \\
\hline
\end{tabular}
\end{center}
\end{table*}

\section{Observational data}
The spectra were obtained with the echelle spectrograph PFES (Panchuk et al., 1998)  
at the prime focus of the 6\,m telescope of SAO RAS in 1997 (the
dates of observations are listed in the last column of Table\,1). The CCD used has
$1160\times1040$ pixels, $16\times16$ microns each. The reduction of the
echelle images was performed in the system ESO MIDAS. When taking positional
measurements of the line profiles, a computer fit of their correct images to
mirror ones was applied.

Eighteen orders spanning a wavelength interval from 4790 to 7520\,\AA\AA\, are
presented in the atlas. The mean spectral resolution limit is 0.5\,\AA. A
thorium-argon hollow-cathode lamp was employed as a comparison spectrum. The
dispersion curves are checked and corrected using the telluric lines O$_2$ and
H$_2$O, which are well visible in Fig.\,3. However, the remaining systematic
error of the presented radial velocities may reach 2--3\,km/s.

\section{Spectroscopic specific characteristics of the objects}

MWC\,314 possesses all of the features of the B[e] supergiant (Lamers et al., 1998),
but for the strong infrared excess. Along with the numerous permitted
low-excitation emission lines, FeII, CrII, ScII and others, there are a few
forbidden lines ([FeII], [CaII] and others) as well as strong emission
H and HeI lines in the spectrum described in detail by Miroshnichenko et al.
(1998). The lines HeI 5876\,\AA\, and 6678\,\AA\, and, possibly, H$_{\beta}$
show P\,Cyg type profiles. The absorption component of the spectrum in the region
accessible to us is represented by weak lines  NII, AlIII, SII, NeI and
others, which are typical of early subclasses of spectral class B.

By the likeness of the profiles and closeness of the radial velocities,
the lines can be divided into several groups. The mean heliocentric velocity
values for these groups are tabulated in Table\,2 (the number of the lines
used is indicated in column 2), while the typical profiles are shown in Fig.\,1.

\begin{table*}[hbtp]
\begin{center}
\caption{Heliocentric radial velocities (km/s) for groups of lines in the
         spectra of MWC\,314  22.11.97.}
\medskip
\begin{tabular}{lrc}
\hline
\hspace*{1cm} Group of lines         &  n &  Vr  \\
\hline
\hspace*{1cm} Emissions:             &    &      \\
 FeII, CrII, ScII, MgI ($\rm r < 1.5$)& 30 &  40  \\
NaI                                  &  2 &  37: \\
FeII ($\rm r > 1.7$); [FeII], [CaII] &  7 &  33  \\
HeI                                  &  4 &  30: \\
SiII                                 &  3 &  16  \\
$\rm H_{\beta}$                      &    &  47  \\
\hspace*{1cm} Absorptions:           &    &      \\
SII, NII, NeI                        & 13 &  81  \\
HeI                                  &  2 &$>$ -800:, -350\\
NaI (I.S.)                           &  2 &  13  \\
DIB                                  & 16 &   3  \\
\hline
\end{tabular}
\end{center}
\end{table*}

\begin{figure}
\includegraphics[angle=0,width=0.6\textwidth,bb=50 100 550 790,clip]{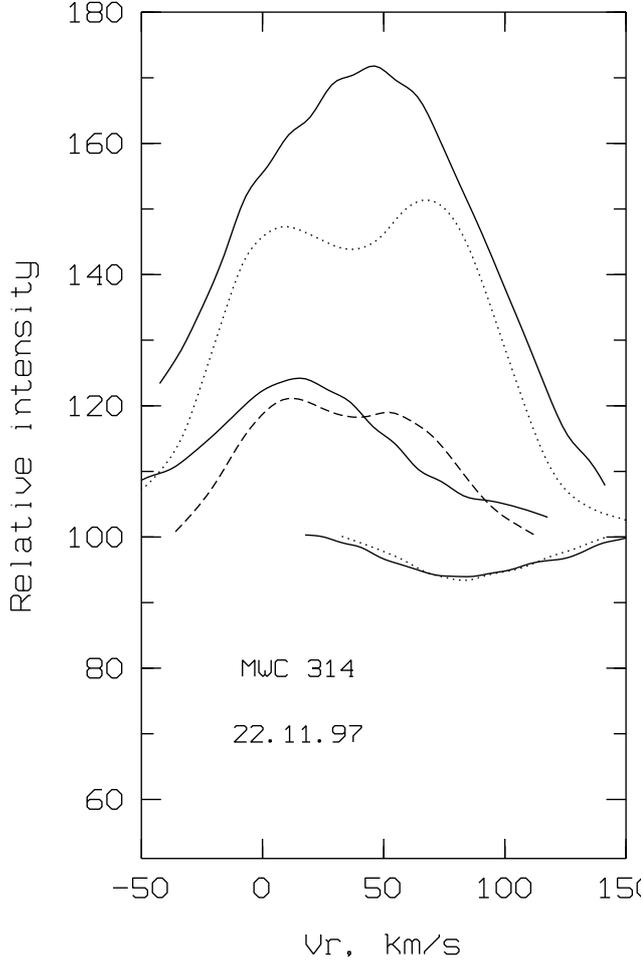}  
\caption{Typical line profiles in the spectrum of MWC\,314. From top to bottom:
emissions FeII(42) 4923, FeII(40) 6432 (dotted line), SiII(2) 6347, [FeII]14F
7155 (dashed line); absorption NeI(1) 6142 and 6402 (dotted line).}
\end{figure}

Apparently, all metallic emission lines are two-peaked. The radial velocities
measured from the profile as a whole are presented in the 1st and 3d lines of
Table\,2. However, more than half of the lines are intensive enough and not
distorted by blending too much to allow radial velocity measurements
from the blue and red peaks separately. Their values for the emission lines
 of moderate intensity are close to 2 and 72 km/s, respectively. The blue peak
is more stable: as can be seen from Fig.\,1, it preserves its position when
going both to stronger permitted lines and to weaker forbidden lines. The red
component in the lines [FeII] is displaced to 52\,km/s, while the displacement
in the strongest FeII lines is as large as 45\,km/s, which causes shift of the line
as a whole. In the emission line SiII, it is likely to merge (at least, with
our resolution) with the blue component, or gets lost on the flat red wing.

The weakest emission  lines of HeI show radial velocities of about 20\,km/s; in
stronger ones, they reach the value indicated in Table\,2, 30\,km/s. The
velocity from H$_{\beta}$ is still higher. This may result from strengthening
of the wind absorption components in these lines.

As regards the photospheric absorption lines, they are displaced redward with
respect to the emission lines, however, no mutual shifts falling outside the
error limits are detected inside this group.

The spectrum of IRC+10420 demonstrates a still wider variety of the line profile
shapes and their differential shifts. The transition from absorptions to
emissions in it is represented by numerous lines with inverse P\,Cyg profiles
(Fig.\,2).

\begin{figure}
\includegraphics[angle=0,width=0.6\textwidth,bb=50 110 560 790,clip]{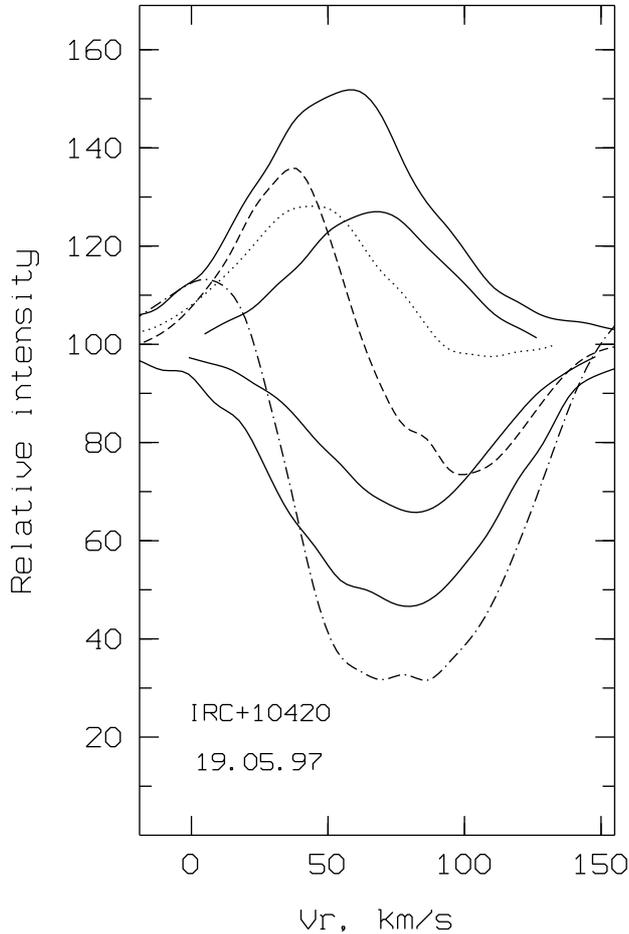}  
\caption{Typical line profiles in the spectrum of IRC\,+10420. From top to bottom:
emissions FeII(40) 6432 and [FeII]14F 7155; emission-absorptions FeII(74) 6456
(dashed line),
FeII(42) 5169 (dashed-dotted line) and FeII(74) 6416 (dotted line); absorptions
NI(3) 7468 and SiII(2) 6347.}
\end{figure}

For the forbidden lines, the radial velocity is practically independent of
line intensity: 64\,km/s from weak [FeII] and 66\,km/s from strong [CaII]. The
value given in the 1st line of Table\,3 is obtained also from measuring the
lower parts of the hydrogen emission line profiles. Weak absorption lines of
HeI and some others formed in the deepest and accessible to us layers of the
atmospheres of the stars also yield a close value (4th line of Table\,3).

\begin{table*}[hbtp]
\begin{center}
\caption{Heliocentric radial velocities (km/s) for groups of lines in
        the spectra of IRC\,+10420 19.05.97.}
\medskip
\begin{tabular}{lrr}
\hline
\hspace*{1cm} Group of lines                      &  n &  Vr  \\
\hline
\hspace*{1cm} Emissions:                          &    &      \\
$[$FeII], [CaII], [OI]                              &  8 &  65  \\
FeII(40.46), MnII                                 & 12 &  54  \\
FeII (48, 49, 74) ($\rm r_{abs} > 0.8$)           &  7 &  40  \\
\hspace*{1cm} Absorptions:                        &    &      \\
HeI, SiII(4,5), OI                                &  4 &  67  \\
FeII (42, 48, 49) ($\rm r < 0.5$)                 &  5 &  70  \\
$\rm H_{\alpha, \beta}$                           &  2 &  71  \\
SiII(2)                                           &  2 &  76  \\
NI                                                &  3 &  80  \\
FeII(74), TiII, CrII, ScII ($\rm r > 0.75$)       & 10 &  90  \\
NaI                                               &  2 &  90  \\
NaI (I.S.)                                        &  2 &  11  \\
DIB                                               & 15 &   2  \\
\hline
\end{tabular}
\end{center}
\end{table*}

Stronger lines free from obvious distortions by emission details (SiII, NI) are
shifted redward with respect to the forbidden lines. This, certainly, refers
to the absorption components of the inverse P\,Cyg profiles in which the shift
is proportional to the relative intensity. For the deepest depressions, the
velocity will be equal to that from the absorption components of the hydrogen
lines (5th and 6th lines of Table\,3), whereas for the shallowest of them
it will reach 90\,km/s (9th line of Table\,3). The blueward shift of the
emission lines is greater the deeper are the absorptions accompanying them.
In the 3rd line of Table\,3 is given the mean velocity value, measured from the
emission peak, for the group of lines with the weakest absorptions. As they
get more intensive, the velocity drops from 40 to 10\,km/s. The shift is,
however, preserved also in the lines that look like pure emission lines
(2nd line of Table\,3).

It is unclear to which group of lines the circumstellar components of the
doublet NaI(1) belong. Possibly these are extension of the series of pure
absorption lines arranged in the order of increasing velocity: SiII(2) ---
76\,km/s, NI(3) --- 80\,km/s, NaI(1) --- 90\,km/s. However, it is not
improbable that we see the absorption components of the inverse P\,Cyg profiles
whose emission components are overlapped by interstellar absorption lines.

Spectroscopy of IRC\,+10420 suggests that envelope surrounding the object
is inhomogeneous and non-stationary. The multilayer structure and, possibly,
the non-sphericity of the envelope, apart from the existence of different
shapes of the profiles shows up also in the deformation of broad absorption
lines by local depressions. Some of them can be seen in Fig.\,2.

The variations of line intensities and profiles of the envelope from 1992
to 1997 (from the spectra taken with the 6\,m telescope of SAO and from
Oudmaijer's (1995) data) were not of systematic character. They manifest themselves 
most clearly in the lines
with inverse P\,Cyg profiles: the general gradient in intensity would change,
i.e. both emission and absorption would be either intensified or alternated
at a time. Apparently, the evolutionarily significant variation of the spectra,
which reflects the elevation of the photosphere (or pseudophotosphere)
temperature, continued: the line HeI 5876\,\AA\ intensified, traces of the
absorption lines HeI 4921\,\AA\ and 5015\,\AA\ appeared.

Note that there are rather deep and sharp absorption lines in the spectrum of
IRC\,+10420 at $\lambda\lambda$ 5693, 6282, 6287 and 6289\,\AA\AA. Their nature
has so far not been established.

The interstellar lines and diffusion bands are strong in the spectra of the two
stars. Their equivalent widths and radial velocities (the last lines of Tables
2 and 3) are indicative of the great distance of both objects. However,
taking account of the interstellar extinction, even adopting the limiting
luminosity, it is difficult to estimate distances at which the observed
stellar velocities would be accounted for by the Galactic rotation alone.
For the longitude in question, the heliocentric radial velocities of disk objects
do not exceed 50--55\,km/s (Avedisova, 1996). This maximum value corresponds
to a distance of about 7\,kpc. Meanwhile, a distance of 2.5\,kpc for MWC\,314
requires M$_V\approx -8^m$, and for IRC\,+10420 even $-9^m$.

At the same time,
it has to be seen which of the values presented in Tables\,2 and 3 could be
thought to be the velocities of the centres of mass of the stars. In the case
of MWC\,314, the most appropriate velocity seems to be the one found from the
two-peaked emission lines of moderate intensity, which, possibly, form in the
gaseous ring surrounding the star. This velocity does not go over the limit
indicated above, but it disagrees with the velocity given by photospheric
absorption lines. There is no such a discrepancy in IRC\,+10420, but the
velocity turns out to run over the limit.

To solve the above-mentioned matters, further observations, preferably with a
higher spectral and temporal resolution, are needed.

\section{The atlas and list of the identified lines}
In Fig.\,3 the spectra of the objects are presented graphically as
relationships between relative intensity and wavelengths in the interval
4790--7520\,\AA\AA. Its fragments correspond to spectral orders. There are
little gaps only between the last three orders. The spectra are displaced
along the horizontal axis so as its marking should correspond to the
laboratory wavelengths for the groups of lines showing the radial velocity
which correspond assumingly to the centre  of mass of the star. For MWC\,314
these are the emission lines of FeII and others (1st line of Table\,2), for
IRC\,+10420 the emission lines [FeII] and others and absorption lines of
HeI and others (1st and 4th lines of Table\,3, respectively).

In the atlas are shown the strongest lines which are present in the spectra of
both stars and the diffusion interstellar bands. For IRC\,+10420, the dots
mark defects.

Table\,4 contains the identification of the lines (column 1), their laboratory
wavelengths used in the estimation of radial velocities (column 2), re
intensities of the profile extrema (columns 3, 5), and radial velocities of
individual lines or their components in km/s (columns 4, 6).

A total of 320 features of stellar (possibly, circumstellar) and interstellar
nature were identified. The telluric lines were not identified. The individual
lines or blends are separated by the skips of the lines.

\newpage
\footnotesize
\begin{center}
\tablecaption{Depths (r) and  values of heliocentric radial velocity
              (${\rm V_r}$) measured for individual lines in spectra of
              stars studied}
\begin{supertabular}{ll|lr|lr}
\hline
\tabletail{&&&&& \\ \hline}
 Ident& $\lambda$, \AA\  &  \multicolumn{2}{c|}{MWC\,314} & \multicolumn{2}{c}{IRC\,+10420} \\
\hline
         &         &       r  &  Vr   &          r    &  Vr     \\
\hline
\quad 1 & 2 &\quad  3  & 4 & \quad 5 &6 \\
\hline
\tablehead{\hline \qquad 1&2&\quad  3&4 &5&6 \\ \hline}
${\rm [FeII] }$ 4F & 4798.3    &    1.06  &       &         1.11: &         \\
TiII (17) & 4798.53          &          &       &               &         \\
&&&&&                                                          \\
TiII (92) &   4805.09 &     1.12 &    43:&          0.75:&         \\
 &  &&&&                                                      \\
 CrII (30)&   4812.34  &    1.07: &   40: &          0.87:&         \\
${\rm [FeII] }$ 20F& 4814.53  &    1.06: &       &          1.09:&         \\
 SII  (9) &    4815.52  &    0.90  &   73  &               &        \\
 &&&&&                                                                      \\
 CrII (30)&    4824.14  &    1.20  &   41  &          0.64 &    78: \\
 FeII (30)&    4825.72  &    1.08: &       &          1.06:&        \\
 &&&&&                                      \\
 FeII (30)&    4833.20  &    1.07  &   42: &          1.06:&        \\
                  &&&&&                                      \\
 CrII (30)&    4836.22  &    1.08  &   43  &          1.06:&    25: \\
          &           &          &       &          0.90:&    92: \\
 &&&&&                                      \\
 FeII (30)&    4840.00  &    1.04: &   42: &               &         \\
                              &&&&&                      \\
 CrII (30)&    4848.25  &    1.15  &   39  &          1.09 &    25:  \\
          &           &          &       &          0.66 &    95:  \\
 &&&&&                                      \\
 CrII (30)&    4856.19  &          &       &          0.88 &    86:  \\
                    &&&&&                                      \\
 ${\rm H_{\beta}}$    &  4861.33 &    8.2  &   47 & 2.16 &     6   \\
                      &          &       &        & 0.47 &    72   \\
          &           &          &       &          1.34 &   145: \\
 &&&&&                                      \\
 CrII (30)&    4864.32  &          &       &          0.68 &    80  \\
 &&&&&                                      \\
 FeII (25)&    4871.27  &    1.04  &   41  &          0.91:&        \\
 &&&&&                                      \\
TiII (114)&   4874.02   &   1.03:  &       &         0.82  &   80:  \\
 &&&&&                                      \\
CrII (30) &   4876.40   &   1.06:  &       &         1.11: &   35:   \\
          &           &          &       &          0.64 &    88   \\
 &&&&&                                      \\
 DIB      &    4881.83  &          &       &               &         \\
 &&&&&                                      \\
 CrII (30)&    4884.60  &    1.03  &       &               &         \\
 SII  (15)&    4885.60  &    0.93: &       &               &         \\
 &&&&&                                      \\
${\rm [FeII] }$ 4F &    4889.62  &    1.05: &   25: &          1.15:&    55: \\
 &&&&&                                      \\
 FeII (36)&    4893.81  &    1.07  &   35: &          1.13:&    55: \\
 NII  (1) &    4895.11  &    0.95  &   78: &               &        \\
 &&&&&                                      \\
 YII  (22)&  4900.12  &          &       &          0.85 &    90: \\
 &&&&&                                                             \\
${\rm [FeII]}$ 20F & 4905.34  &    1.04  &   21: &        &        \\
 &&&&&                                      \\
TiII (114)&   4911.19   &   1.07   &  37   &         1.08  &    0:   \\
          &           &          &       &          0.85 &    70:  \\
 &&&&&                                      \\
 HeI  (48)&   4921.93   &   1.28:  &  15:  &         0.94: &         \\
 FeII (42)&   4923.92   &   1.72   &  39   &         0.40  &   78    \\
 &&&&&                                      \\
FeI  (318)&   4957.5:   &   1.03   &       &               &         \\
 &&&&&                                      \\
 DIB      &   4963.96   &   0.96   &  10:  &               &        \\
 &&&&&                                      \\
${\rm [TiII] }$ 23F&   4982.73?  &   1.04   &       &         1.07: &        \\
 &&&&&                                      \\
 NII  (24)&   4987.37   &   0.97:  &  77:  &               &        \\
 &&&&&                                      \\
FeII (25) &   4991.11   &   1.07:  &       &         1.14  &   88   \\
FeII (36) &   4993.35   &   1.26   &  45:  &         1.25  &   50:  \\
&&&&& \\
FeII (25) & 5000.73   &   1.20   &  37   &         1.03: &        \\
 &&&&&                                      \\
NII  (4)  &   5002.70   &   0.92   &  77:  &               &         \\
 &&&&&                                      \\
NII  (24) &   5007.33   &   0.88   &       &               &         \\
SII  (7)  &   5009.54   &   0.88   &  69:  &               &         \\
 &&&&&                                      \\
HeI  (4)  &   5015.68   &   1.77   &  25   &         0.88  &         \\
FeII (42) &   5018.44   &   1.90   &  32   &         0.42  &   72    \\
 &&&&&                                      \\
 SII  (1) &   5027.19   &   0.97   &  79   &               &        \\
 &&&&&                                      \\
 ScII (23)&   5031.02   &   1.06   &  37:  &         0.77  &   80   \\
 SII  (7) &   5032.41   &   0.91   &  81   &               &        \\
 &&&&&                                      \\
 FeII (36)&   5036.93   &   1.05   &       &         1.06  &   75:  \\
 &&&&&                                      \\
 SiII (5) &    5041.03  &    1.07  &   16: &          0.92 &        \\
 &&&&&                                      \\
 NII  (4) &    5045.10  &    0.94  &   83: &               &         \\
 HeI  (47)&    5047.74  &          &       &               &         \\
 &&&&&                                      \\
 SiII (5) &    5056.00  &    1.12  &   16: &          0.80 &    68   \\
 &&&&&                                      \\
 &&&&&                                      \\
TiII (113)&    5072.29  &    1.09  &   44: &          1.08:&    35:  \\
          &           &          &       &          0.89:&    95:  \\
 NII  (10)&   5073.59   &   0.98   &  82:  &               &         \\
 &&&&&                                      \\
 YII  (20)&   5087.42?  &          &       &         0.92: &        \\
 &&&&&                                      \\
 CrII (24)&   5097.32   &   1.02:  &       &               &        \\
 &&&&&                                      \\
 FeII (35)& 5100.65   &   1.12   &  40   &         1.13  &   43   \\
 SII  (7) & 5103.30   &   0.93   &  79   &               &        \\
 &&&&&                                      \\
FeI(16,36)&  5107.5:    &          &       &        1.08:  &         \\
 &&&&&                                      \\
${\rm [FeII] }$ 19F&5111.63   &   1.05:  &  14:  &               &         \\
 &&&&&                                      \\
 FeII (35)& 5120.34   &   1.08   &  38:  &         1.08  &   30    \\
 &&&&&                                      \\
 TiII (86)&   5129.14   &   1.15   &  41:  &         1.10: &    2:   \\
          &           &          &       &         0.85  &    75:   \\
 &&&&&                                      \\
 FeII (35)&   5132.66   &   1.20   &  38   &         1.18  &   37:  \\
 &&&&&                                      \\
 FeII (35)&   5136.79   &   1.09   &  40   &         1.08  &   38   \\
&&&&&                                      \\
 FeII (35)&   5146.11   &   1.11   &  39   &         1.13  &   44   \\
 &&&&&                                      \\
 FeII     &   5149.46?  &          &       &         0.93  &        \\
&&&&&                                      \\
 FeI  (16)&   5150.84   &          &       &         1.05: &   50:  \\
&&&&&                                      \\
 TiII (70)&   5154.07   &   1.15   &  41   &         1.15  &   18    \\
          &           &          &       &           0.73&    78   \\
      &&&&&                                                             \\
${\rm [FeII] }$ 19F&   5158.78   &   1.16   &  34:  &         1.22  &   60    \\
 &&&&&                                      \\
 FeII (35)&   5161.18   &   1.06   &  39:  &         1.04: &   58:   \\
 &&&&&                                      \\
 MgI  (2) &   5167.33   &          &       &         1.12  &   30:   \\
 FeII (42)&   5169.03   &   1.68   &  33   &         1.10  &   10:  \\
          &             &          &       &         0.32  &    82   \\
&&&&&                                                             \\
 FeII (35)&   5171.60   &          &       &         1.18: &   65:  \\
 MgI  (2) &   5172.68  &    1.30: &   37: &          0.86:&    80:  \\
 &&&&&                                      \\
 MgI  (2) &    5183.61  &    1.24  &   43  &          1.12 &    32  \\
          &           &          &       &           0.78&     86 \\
 TiII (86)&    5185.90  &          &       &          1.15 &    29  \\
          &           &          &       &          0.78 &    88   \\
 TiII (70)&   5188.68   &   1.10   &  45:  &         1.08: &   20    \\
          &           &          &       &           0.55&     86  \\
 &&&&&                                      \\
 FeII (49)&   5197.57   &   1.46   &  40   &         1.28  &   27    \\
          &           &          &       &           0.52&     92  \\
 &&&&&                                      \\
 SII  (39)& 5201.15:  &   0.93   &  84:  &               &         \\
 &&&&&                                      \\
TiII (103)&   5211.54   &   1.09   &       &         1.14  &   18   \\
          &           &          &       &           0.93&:    78:\\
 SII  (39)&   5212.61   &   0.94   &  81:  &               &        \\
 &&&&&                                      \\
${\rm [FeII] }$ 19F&   5220.06?  &          &       &    1.07: &   59:  \\
&&&&&                                                                  \\
 TiII (70)&   5226.54   &   1.20   &  42:  &         1.10  &   14:  \\
          &           &          &       &          0.70 &    72: \\
 &&&&&                                      \\
 FeII (49)&    5234.62  &    1.48  &   25  &          1.15:&     5:  \\
          &           &          &       &           0.60&     72  \\
 CrII (43)&   5237.32   &   1.17:  &  41:  &         1.02: &    8:   \\
          &           &          &       &          0.72 &    73   \\
 ScII (26)&   5239.82   &   1.04:  &  42:  &         1.02: &   18:   \\
          &           &          &       &          0.90 &    72:  \\
&&&&&                                      \\
 CrII (23)&   5246.76   &   1.05   &  45:  &         1.03: &   25:   \\
          &           &          &       &          0.96:&    90:  \\
 &&&&&                                      \\
 CrII (23)&   5249.43   &   1.04:  &       &         1.03: &   27:  \\
 &&&&&                                      \\
 FeII (49)&   5254.93   &   1.26   &  46:  &         1.16  &   32   \\
 FeII (41)&   5256.93   &   1.1:   &  37:  &         1.10  &   40:   \\
 &&&&&                                      \\
 FeII     &   5260.26?  &          &       &         0.91  &        \\
 &&&&&                                                                      \\
${\rm [FeII] }$ 19F&  5261.62   &   1.12   &       &         1.21  &   62   \\
 TiII (70)&   5262.10   &          &       &               &         \\
 &&&&&                                                                       \\
 FeII (48)&   5264.80   &   1.20   &  42   &         1.14  &   28    \\
 &&&                                     &         0.83  &   90    \\
TiII (103)&   5268.63   &   1.20   &       &         1.10: &         \\
FeI  (15) &   5269.54  &          &       &               &         \\
 &&&&&                                                                       \\
${\rm [FeII] }$ 18F& 5273.36   &   1.10:  &       &               &         \\
 CrII (49)& 5274.99   &          &       &               &        \\
 FeII (49)& 5276.00   &   1.52   &       &         1.13  &        \\
          &           &          &       &          0.45 &    83  \\
 &&&&&                                                                      \\
 CrII (43)& 5279.88   &   1.08   &  41:  &         1.09  &   32:  \\
 &&&                                     &         0.91  &   87: \\
 CrII (43)& 5280.08   &          &       &               &        \\
 &&&&&                                                                      \\
 &&&&&                                                                      \\
 FeII (41)& 5284.10   &   1.35   &  42   &         1.29  &   35   \\
          &           &          &       &          0.84 &    88   \\
          &           &          &       &               &         \\
 FeII     & 5291.67  &          &       &          0.96:&    52:  \\
 &&&&&                                                                       \\
 CrII (24)& 5305.85   &   1.09   &  50:  &         1.11  &   30    \\
          &           &          &       &           0.94&:    92: \\
 &&&&&                                                                       \\
 CrII (43)&   5308.42   &   1.06   &  48:  &         1.12  &   30    \\
          &           &          &       &          0.95:&    87: \\
 &&&&&                                                                      \\
 CrII (43)&   5310.69   &          &       &         1.06  &   31   \\
          &           &          &       &           0.98&:    82:\\
           &&&&&    \\
 CrII (43)&   5313.58   &   1.08:  &       &         1.08: &   20:  \\
          &           &          &       &        0.86   &  82    \\
FeII(49,48)&   5316.65: &    1.82  &   32  &          1.26 &    16  \\
          &           &          &       &        0.43   &  83    \\
 &&&&&                        \\
 SII  (38)&  5320.70    &  0.92    & 80:   &               &         \\
 &&&&&                                                                       \\
 FeII (49)&  5325.56    &  1.25    & 40    &        1.21   &  28     \\
 &&&                                     &        0.88   &   85    \\
 FeI  (15)&  5328.04    &  1.12    &       &        1.18   &         \\
 FeI  (37)&  5328.53    &          &       &               &         \\
          &           &          &       &               &         \\
${\rm [FeII] }$ 19F&   5333.65   &          &     &  1.18: &         \\
 CrII (43)&   5334.86   &   1.12   &  39:  &               &        \\
          &           &          &       &        0.87   &  72    \\
 TiII (69)&   5336.79   &   1.20:  &  43:  &         1.18  &        \\
 FeII (48)&   5337.73   &          &       &               &        \\
 &&&&&                                                                  \\
 FeII (49)&   5346.56   &   1.04   &       &         1.05: &            \\
 &&&&&                                                                  \\
 FeII (48)&   5362.86   &   1.43   &  35   &         1.22  &   28   \\
          &           &          &       &         0.75  &   87   \\
 &&&&&                                                                  \\
 CrII (29)&   5369.30?  &          &       &         1.03: &         \\
 FeI  (15)&   5371.49?  &   1.04:  &       &         1.05: &         \\
 &&&&&                                                                   \\
${\rm [FeII] }$ 19F&   5376.47   &   1.04   &  24:  &         1.07: &   70:   \\
 &&&&&                                                                   \\
 TiII (69)&   5381.02   &   1.08   &  43   &         1.16  &   36    \\
          &           &          &       &          0.93 &    95   \\
 &&&&&                                                                   \\
 TiII (80)&   5396.30   &          &       &               &        \\
 FeI  (15)&   5397.13   &   1.07:  &       &         1.13  &   35   \\
 &&&&&                                                                      \\
 DIB      & 5404.5:   &   0.95   &   7:  &         0.96: &    2:  \\
 &&&&&                                                                      \\
 FeI  (15)&   5405.78   &   1.05:  &       &         1.03  &   56:  \\
 CrII (23)&   5407.61   &   1.10:  &  45:  &         1.07  &   40:  \\
 &&&&& \\
${\rm [FeII] }$ 17F&   5412.65   &          &       &               &        \\
 FeII (48)&   5414.07   &   1.18   &  41   &         1.16  &   45:  \\
 &&&&&                                                                       \\
 TiII (69)&   5418.78   &          &       &         1.10  &   44    \\
          &           &          &       &         0.91  &   90    \\
 CrII (23)&   5420.92   &   1.12   &  39:  &         1.10: &   45:   \\
 &&&&&                                                                       \\
 FeII (49)&   5425.25   &   1.32   &  42   &         1.18  &   40    \\
          &           &          &       &          0.92 &   100   \\
&&&&&                                      \\
 SII  (6) &   5428.64   &   0.98:  &  88:  &               &         \\
 &&&&&                                                                      \\
 FeII (55)&   5432.98   &   1.16   &  39   &         1.12  &   50:  \\
 &&&&&                                                                      \\
 FeI  (15)&   5446.91   &   1.02:  &       &         1.09  &   48   \\
 &&&&&                                                                      \\
 SII  (6) &   5453.81   &   0.90   &  81   &               &        \\
 FeI  (15)&   5455.61   &   1.05   &       &         1.06: &   48:  \\
 &&&&&                                                                      \\
 FeII     &   5466.92?  &   1.03:  &       &               &        \\
 &&&&&              \\
 CrII (50)&   5472.62   &   1.02:  &  43:  &               &         \\
 SII  (6) &   5473.59   &   0.94   &  83:  &               &         \\
 &&&&&                                                                        \\
 CrII (50)&   5477.45   &          &       &               &          \\
 CrII (50)&    5478.36  &    1.10  &   36: &          1.12 &    35:   \\
          &           &          &       &         0.96  &   90:    \\
 &&&&&                                                                        \\
 DIB      &   5487.45   &   0.97:  &       &               &          \\
 &&&&&                                                                        \\
 TiII (68)&   5490.65   &   1.06   &       &         1.07  &   55:    \\
 &&&&&                                                                       \\
 FeI  (15)&   5497.52   &   1.04:  & 36:   &         1.03: &   60:   \\
 &&&&&                                                                       \\
 CrII (50)& 5502.08   &   1.12:  &       &         1.12: &         \\
 CrII (50)&   5503.20   &          &       &               &         \\
 &&&&&                                                                       \\
 FeII     &   5506.20   &          &       &         0.95: &   55:   \\
 CrII (50)&    5508.62  &    1.10: &       &          1.06 &    43   \\
 CrII (23)&   5510.70   &   1.04:  &       &         1.08  &   39    \\
          &           &          &       &         0.98: &  108:    \\
 &&&&&                                                                        \\
 FeII (56)&   5525.11   &   1.08   &       &         1.10: &   66:     \\
 ScII (31)&   5526.81   &   1.15   & 41    &         1.12: &           \\
          &           &          &       &         0.78  &   80      \\
 &&&&&                                                                         \\
 TiII (68)&   5529.94   &   1.08   & 39    &         1.09  &   72      \\
 &&&&&                                                                         \\
 FeII (55)&   5534.83   &   1.44   & 40    &         1.28  &   39      \\
          &           &          &       &         0.81  &   97      \\
 &&&&&                                                                        \\
 DIB      &   5544.97   &   0.97:  & -3:   &         0.96: &    3:    \\
 &&&&&                                                                        \\
 SII  (6) &   5556.01   &   0.98   & 79:   &               &          \\
 &&&&&                                                                        \\
 SII  (6) &    5564.96  &    0.95  &  79   &               &          \\
 &&&&&                                                                        \\
 FeII     &    5567.84  &    1.05  &       &          1.07 &    63    \\
 &&&&&                                                                        \\
${\rm [OI] }$ 3F   & 5577.34  &          &       &               &          \\
 &&&&&                                                                        \\
 SII  (11)&    5578.85  &    0.98  &  81:  &               &           \\
 &&&&&                                                                         \\
 FeI (686)&   5586.76   &   1.07   &       &         1.08  &   60:     \\
 &&&&&                                                                         \\
FeII (55) &   5591.37   &   1.05   & 38    &         1.04  &   65:     \\
 &&&&&                                                                         \\
SII  (11) & 5606.11   &   0.95   & 78:   &               &           \\
 &&&&&                                                                         \\
DIB       &   5609.96   &   0.98   & -3:   &         0.97  &    8:    \\
 &&&&&                                                                        \\
FeI  (686)&   5615.64   &   1.03   &       &         1.04  &   45:    \\
SII  (11) &   5616.63   &   0.98   & 76:   &               &          \\
 &&&&&                                                                        \\
FeI  (686)&   5624.54   &   1.02:  &       &         1.02: &          \\
 &&&&&                                                                        \\
FeII (57) &   5627.49   &   1.07   & 43    &         1.09  &   55:    \\
 &&&&&                                                                        \\
SII(14,11)&   5640.15:  &   0.93   & 77:   &               &          \\
 &&&&&                                                                        \\
 ScII (29)&   5640.98   &          &       &         1.04  &   36      \\
          &           &          &       &         0.95  &   90      \\
 &&&&&                                                                         \\
 SII  (6) &   5645.62   &   0.98:  & 83:   &               &           \\
 SII  (14)&   5646.98   &   0.93   & 81    &               &           \\
&&&&&                                      \\
 ScII (29)&   5657.87   &   1.16   & 49    &         1.12  &   36      \\
          &           &          &       &         0.86  &   95      \\
 SII  (11)&   5659.95   &   0.95:  & 90:   &               &           \\
 &&&&&                                                                        \\
 YII  (38)&   5662.94   &          &       &         1.06  &   37     \\
          &           &          &       &         0.94  &   93     \\
 SII  (11)&   5664.73   &   0.95:  & 81:   &               &          \\
 NII  (3) &   5666.63   &   0.95   & 77:   &               &          \\
 ScII (29)&   5667.16   &   1.03:  &       &         1.04  &   35     \\
          &           &          &       &         0.98: &   86:    \\
&&&&& \\
 ScII (29)&   5669.03   &   1.02:  & 38:   &         1.04  &   36     \\
          &           &          &       &          0.95 &    88    \\
 &&&&&     \\
 NII  (3) &   5676.02   &   0.95   & 85:   &               &           \\
CrII (189)&   5678.42   &   1.04:  &       &         1.03: &   37:     \\
          &           &          &       &         0.98: &  100:      \\
 NII  (3) &   5679.56   &   0.91   & 89:   &               &            \\
 &&&&&                                                                          \\
 ScII (29)&   5684.19   &   1.07   & 43    &         1.08  &   42       \\
          &           &          &       &          0.96 &    94:     \\
 NII  (3) &   5686.21   &   0.97   & 76    &               &            \\
 &&&&&                                                                          \\
          &   5687?     &          &       &         0.92  &            \\
 &&&&&                                                                         \\
 NaI  (6) &   5688.21?  &          &       &         0.98  &   79:     \\
 &&&&&                                                                         \\
          &   5692.5?   &          &       &         0.72  &           \\
 &&&&&                                                                         \\
 AlIII(2) &   5696.60   &   0.93   & 83:   &               &           \\
 &&&&&                                                                         \\
          & 5700.0?     &          &       &         0.91  &           \\
 &&&&&                                                                         \\
 DIB      &    5705.1   &    0.94  &   6   &          0.96 &     5:    \\
 &&&&&                                                                         \\
 NII  (3) &    5710.77  &    0.96  &  83   &               &            \\
 &&&&&                                                                          \\
 DIB      &    5719.43  &    0.98: &   8:  &          0.98:&     4:     \\
 &&&&&                                                                          \\
 AlIII(2) &    5722.73  &    0.96  &  87:  &               &            \\
 &&&&&                                                                          \\
 SiIII(4) &    5739.73  &    0.95: &  83:  &               &            \\
 &&&&&                                                                          \\
${\rm [NII] }$ 3F  &5754.8   &          &       &               &           \\
 &&&&&                                                                         \\
 DIB      &    5766.25  &    0.96  &  -6:  &          0.96 &     0     \\
 &&&&&                                                                         \\
 DIB      &    5769.1   &    0.98: &   4:  &               &           \\
 &&&&&                                                                         \\
 DIB      &    5772.6   &    0.98  &   6:  &          0.98:&    10:    \\
 &&&&&                                                                         \\
 DIB      &    5776.08  &    0.97  &       &          0.98:&           \\
 &&&&&                                                         \\
 DIB      &    5780.41  &    0.67  &   7   &          0.74 &     4     \\
 &&&&&                                                                         \\
 DIB      &    5784.9   &    0.97  &   1:  &          0.97:&     4:     \\
 &&&&&                                                                          \\
 DIB      &    5789.06  &    0.98  &       &               &            \\
 &&&&&                                                                          \\
 DIB      &    5797.03  &    0.82  &   2:  &          0.75 &     8      \\
 &&&&&                                                                          \\
 DIB      &  5809.1:  &    0.97  &   4:  &          0.97:&     6:     \\
 &&&&&                                                                          \\
FeII (163)&   5813.67?  &   1.06   & 40:   &         1.04: &   38:     \\
 &&&&&                                                                         \\
 DIB      &   5818.85   &          &       &         0.98: &    3:     \\
 &&&&&                                                                         \\
FeII (182)&   5835.49?  &   1.04   &       &         1.04  &   47:     \\
 &&&&&                                                                         \\
 DIB      &    5844.2   &    0.97  &       &          0.97 &     3     \\
 &&&&&                                                                         \\
 DIB      &    5849.79  &    0.93  &   6   &          0.87 &     2     \\
 &&&&&                                                                  \\
 NeI  (6) &    5852.49  &    0.97  &  85:  &               &           \\
 &&&&&                                                               \\
 HeI  (11)&    5875.72  &    0.85  &-700: &           0.91 &   65:      \\
          &             &          &-450: & &           \\
          &             &          &-350: & &           \\
          &             &    4.00  &   28  &&\\
 &&&&&                                                                          \\
 NaI  (1) &    5889.95  &    1.30  &  37:  &               &            \\
          &           &     0.06 &   13  &           0.09&     10     \\
          &           &          &       &          0.53 &    92:     \\
 &&&&&                                                                          \\
 NaI  (1) &    5895.92  &    1.25  &  36:  &               &            \\
          &           &     0.09 &   12  &           0.10&     12    \\
          &           &          &       &          0.54 &    86:    \\
 &&&&&                                                                         \\
 VII  (98)& 5928.86   &          &       &         1.05: &   68:     \\
 &&&&&                                                                         \\
FeII (182)&   5952.52?  &   1.05:  & 46:   &         1.05  &   50:     \\
 &&&&&                                                                         \\
 SiII (4) &    5957.56  &    0.95  &  75:  &          0.91 &    66     \\
 &&&&&                                                                         \\
 SiII (4) &    5978.93  &    1.08  &  17   &          0.94:&    75:    \\
 &&&&&                                                                         \\
 FeII (46)&    5991.37  &    1.38  &  41   &          1.43 &    55      \\
 &&&&&                                                                          \\
 DIB      &  6004.9   &    0.97  &   6:  &          0.98:&    -1:     \\
 &&&&&                                                                          \\
 DIB      &    6010.58  &    0.95  &  10:  &          0.98:&     5:     \\
 &&&&&                                                                          \\
 DIB      &    6019.34  &    0.98: &   8:  &          0.98:&            \\
 &&&&&                                                                          \\
 VII  (97)&    6028.26  &    1.05  &  37:  &          1.05 &    62:    \\
 VII  (97)&    6031.07  &    1.02: &       &          1.03 &    60:    \\
 &&&&&                                                                         \\
 DIB      &     6037.56 &     0.97 &   -5: &           0.98&:     4:   \\
 &&&&&                                                                         \\
FeII (200)&   6045.46   &   1.02   & 46:   &         1.03  &           \\
 &&&&&                                                                         \\
CrII (105)&   6053.48   &   1.04   & 42:   &         1.05  &   50:     \\
 &&&&&                                                                         \\
 DIB      &   6065.38   &   0.98   &  5:   &         0.99: &          \\
 &&&&&                                                                        \\
 NeI  (3) &   6074.34   &   0.96   & 82:   &               &           \\
 &&&&&                                                                         \\
 FeII (46)&    6084.10  &    1.22  &  41   &          1.25 &    58     \\
 &&&&&                                                                         \\
 DIB      &    6089.75  &    0.96  &   0   &          0.97 &     3     \\
 &&&&&                                                                         \\
 NeI      &    6096.16  &     0.96 &    85:&               &                \\
 &&&&&                                                                         \\
FeII (200)& 6103.54   &   1.07   & 40    &         1.05  &   41:    \\
 &&&&&                                                                        \\
 DIB      &    6113.2   &          &       &          0.98:&          \\
 FeII (46)&    6113.32  &    1.11  &  50:  &          1.14 &    64    \\
 FeII (46)&    6116.05  &    1.05: &       &          1.05 &    65:   \\
          &           &          &       &               &          \\
 MnII (13)&    6122.43? &    1.02: &       &          1.03:&          \\
 &&&&&                                                                        \\
 FeII (46)&    6129.71  &    1.08  &  40   &          1.10 &    60    \\
 &&&&&            \\
FeI (169) &   6136.61   &          &       &               &          \\
 FeI (207)&   6137.69   &   1.02:  &       &         1.08: &          \\
 &&&&&                                                                         \\
 BaII (2) &    6141.72  &          &       &          1.05 &    42:    \\
 &&&&&                                                                         \\
 NeI  (1) &    6143.06  &    0.94  &  83   &               &           \\
 &&&&&                                                                         \\
 FeII (74)&    6147.74  &    1.55  &  41   &          1.26 &    40     \\
 FeII (74)&    6149.25  &    1.35  &       &          1.20 &    46:    \\
          &           &          &       &          0.96 &    98:    \\
 &&&&&                                                                        \\
 OI   (10)&    6156     &          &       &          0.95 &          \\
 OI   (10)&    6158.18  &          &       &          0.95 &    65:   \\
 &&&&&                                                                        \\
 NeI  (5) &    6163.59  &    0.95  &  81   &               &          \\
 &&&&&                                                                        \\
FeII (200)&    6175.15  &    1.05  &  40:  &          1.05:&          \\
 &&&&&                                                                        \\
FeII (163)&    6179.39  &    1.04  &  41:  &          1.05:&    40:   \\
 &&&&&          \\
 FeI (169)&    6191.56  &    1.03  &  40   &          1.06 &    65:   \\
 &&&&&                                                                        \\
 DIB      &    6195.95  &    0.93  &   4:  &          0.92 &     0     \\
 &&&&&                                                                         \\
 DIB      &  6203.06  &    0.87  &   8:  &          0.90 &     2     \\
 &&&&&                                                                         \\
 DIB      &    6212.2   &    0.97: & -10:  &          0.98:&    -5:    \\
 &&&&&                                                                         \\
 FeII (34)&    6219.54  &    1.04  &  47:  &          1.06 &    75:    \\
 &&&&&                                                                         \\
 FeII (34)&   6229.34   &   1.07   & 51:   &         1.10  &   80:    \\
 FeI (207)&   6230.74   &   1.04   & 33:   &         1.05: &   75:    \\
 &&&&&                                                                        \\
 FeII (74)&   6238.39   &   1.48   & 46:   &         1.27  &   42     \\
 FeII (74)&   6239.95   &   1.20:  &       &         1.19  &   58     \\
 &&&&&                                                                        \\
 ScII (28)&   6245.62   &          &       &         1.09  &   40     \\
 FeII (74)&   6247.55   &   1.40   & 41    &         1.28  &   38     \\
          &           &          &       &           0.81&    100   \\
 &&&&&            \\
 FeI (169)&    6252.56  &          &       &          1.03:&    70:   \\
 &&&&&                                                                        \\
          &    6259?    &    0.99: &       &          0.97:&           \\
 &&&&&                                                                         \\
 DIB      &    6269.7  &    0.93  &       &          0.91 &     -1     \\
 FeI(342) &    6270.24  &    1.06  &       &          1.09 &    65:    \\
 &&&&&                                                                         \\
          &    6282.6?  &          &       &          0.85:&           \\
 DIB      &    6283.86  &    0.67  &  10   &          0.74 &     3     \\
          &    6286.6?  &          &       &          0.81 &           \\
          &    6288.5?  &          &       &          0.79 &          \\
 &&&&&                                                                        \\
${\rm [OI] }$ 1F   &   6300.3  &    1.04  &       &           1.21&     66   \\
 &&&&&                                                                        \\
${\rm [SIII] }$ 3F & 6312.06? &   1.10   &  16:  &               &          \\
 &&&&&                                                                        \\
 FeII     &    6317.99  &   1.25   &  41   &               &          \\
 &&&&&                                                                        \\
FeII (199)&    6331.96  &   1.09   &  45   &          1.05 &    45:   \\
 &&&&&                 \\
 NeI  (1) &    6334.43  &   0.97   &  73   &               &          \\
 &&&&&                                                                        \\
 SiII (2) &    6347.10  &   1.23   &  16   &          0.47 &    75     \\
 &&&&&                                                                         \\
 DIB      &    6353.5   &   0.96   &   2:  &          0.98 &    -8:    \\
 &&&&&                                                                         \\
 DIB      &    6362.4   &   0.98   &   3:  &          0.97 &    -5:    \\
 &&&&&                                                                         \\
${\rm [OI] }$ 1F   & 6363.78  &          &       &          1.05 &    67     \\
 &&&&&                                                                         \\
FeII (74) &    6369.47  &  1.38    &  53:  &          1.32 &    56    \\
 SiII (2) &    6371.36  &  1.10:   &       &          0.59 &    78    \\
 &&&&&                                                                        \\
 DIB      &    6376.1   &  0.96    &  12:  &          0.97 &     3:   \\
 &&&&&                                                                        \\
 DIB      &    6379.2   &  0.86    &   1   &          0.85 &     2    \\
 &&&&&                                                                        \\
 FeII     &    6383.72  &  1.18    &  50:  &               &          \\
 FeII     &    6385.45  &  1.1:    &       &          1.05:&          \\
          &           &          &       &               &          \\
 FeI (168)&    6393.61  &  1.02:   &  40:  &          1.05 &    65:   \\
 &&&&&                                                                         \\
 FeI (816)&  6400.01? &  1.02:   &       &          1.03:&    71:    \\
 NeI  (1) &    6402.25  &  0.92    &  83   &               &           \\
 &&&&&                                                                         \\
 FeII (74)&    6407.25  &  1.10    &  39   &          1.10 &    54     \\
 &&&&&                                                                         \\
 FeII (74)&    6416.92  &  1.32    &  41   &          1.28 &    43     \\
          &           &          &       &           0.98:&   108:   \\
 &&&&&                                                                        \\
 DIB      &    6425.4   &  0.96    &   6:  &          0.97 &     9:   \\
 &&&&&                                                                        \\
 FeII (40)&    6432.68  &  1.55    &  39   &          1.51 &    54    \\
 &&&&&                                                                        \\
 DIB      &    6439.4   &  0.97    &   2:  &          0.98 &     5:   \\
${\rm [FeII] }$ 15F&    6440.4   &  1.02:   &  18:  &               &          \\
 FeII     &    6442.95  &  1.05    &  35:  &               &          \\
 &&&&&                                                                        \\
 DIB      &    6445.4   &  0.97    &       &          0.97 &     5:   \\
FeII (199)&    6446.43  &  1.05    &  48:  &          1.04 &    42:   \\
 DIB      &    6449.2   &          &       &          0.97 &     7:    \\
 &&&&&                                                                         \\
 FeII (74)&    6456.38  &  1.60    &  40   &          1.35 &    38     \\
          &           &          &       &          0.74 &   101     \\
 &&&&&                                                                         \\
FeII (199)&    6482.19  &  1.13    &  42   &          1.07 &    30:    \\
          &           &          &       &          0.92 &    95:    \\
 &&&&&                                                                         \\
 TiII (91)&    6491.57  &  1.15    &  43:  &          1.12:&    40:   \\
 FeII     &    6491.67  &          &       &               &          \\
 FeII     &    6493.03  &          &       &               &          \\
 &&&&&                                                                        \\
 NeI  (3) &  6506.53  &  0.96    &  90:  &               &          \\
 &&&&&                                                                        \\
 FeII (40)&    6516.08  &  1.65    &  41   &          1.46 &    50:   \\
 &&&&&                                                                        \\
${\rm  H_{\alpha}}$   &    6562.81  &          &       &          6.2  &     7    \\
          &           &          &       &          1.93 &    70    \\
          &           &          &       &          3.66 &   138    \\
${\rm [NII] }$ 1F  &    6583.6   &          &  10   &               &           \\
 FeII     &    6586.70? &          &  45:  &               &           \\
 &&&&&                                                                         \\
 DIB      &    6597.39  &  0.98:   &       &          0.98:&           \\
 NeI  (6) &    6598.95  &          &  90:  &               &           \\
 &&&&&                                                                         \\
 ScII (19)&  6604.59  &  1.07:   &       &          1.06 &    45:    \\
 TiII (91)&    6606.95  &          &       &          1.05 &    40:    \\
 &&&&&                                                                        \\
 DIB      &    6613.62  &  0.75    &   3   &          0.76 &     0    \\
 &&&&&                                                                        \\
FeII (210)&    6627.25  &  1.07    &  39   &          1.02:&          \\
 &&&&&                                                                        \\
 DIB      &    6660.65  &  0.94    &   5:  &          0.93 &     0    \\
 &&&&&                                                                        \\
 HeI  (46)&    6678.15  &  0.95 & -400:  &        &          \\
          &             &       & -290:  &        &          \\
          &             &       & -115:  &        &          \\
          &           &  2.20    &  22   &               &          \\
TiII (112)&    6680.26? &          &       &          1.05 &          \\
 &&&&&                                                                        \\
 DIB      &    6699.37  &  0.95    &   4:  &          0.95:&     0:    \\
 DIB      &  6701.98  &  0.97    &  -8:  &          0.98:&     6:    \\
          &           &          &       &               &           \\
${\rm [SII] }$ 2F  & 6717.0    & 1.03     & 27:   &               &           \\
TiII (112)&   6717.91   & 1.02     & 35:   &         1.07  &   42:     \\
 &&&&&                                                                         \\
 DIB      &   6740.96   & 0.98:    &       &               &           \\
 &&&&&                                                                         \\
 DIB      &   6770.05   & 0.98:    & 10:   &               &          \\
 &&&&&                                                                        \\
YII  (26) &   6795.41   &          &       &         1.03: &   55:    \\
 &&&&&                                                                        \\
 DIB      &  6811.44  &  0.98    &   7:  &          0.99:&          \\
 &&&&&                                                                        \\
 DIB      &    6827.28  &  0.98    &   2:  &          0.99:&          \\
${\rm [FeII] }$ 31F& 6829.01? &  1.04    &  23:  &          1.06:&          \\
 &&&&&                                                                        \\
 DIB      & 6841.59  &  0.99    &       &          0.99:&          \\
 DIB      &    6843.44  &  0.98    &   0:  &          0.98:&     3:   \\
 &&&&&                                                                         \\
 DIB      &    6852.9   &  0.98    &       &          0.98 &           \\
 &&&&&                                                                         \\
 DIB      &    6860.02  &  0.98:   &       &          0.97 &           \\
 &&&&&                                                                         \\
 DIB      &  6993.18  &  0.83    &   4:  &          0.88 &     8:    \\
 &&&&&                                                                         \\
${\rm [TiII] }$ 17F& 6999.99? &  1.06:   &  18:  &          1.05:&           \\
 &&&&&                                                                        \\
 HeI  (10)&  7065.2   &  0.95:   &       &               &          \\
          &           &   4.7    &   35  &               &          \\
          &           &          &       &               &          \\
 DIB      &  7105.93  &  0.98    &   5:  &               &          \\
 &&&&&                                                                        \\
 DIB      &    7119.94  &  0.97    & -12:  &          0.96 &          \\
 &&&&&                                                                        \\
FeII (197)&    7135.02? &  1.10    &  38   &          1.02:&          \\
          &           &          &       &          0.97:&          \\
 &&&&&                                                                        \\
${\rm [FeII] }$ 14F&    7155.14  &  1.20    &  33:  &          1.27 &     65    \\
 &&&&&                                                                         \\
TiII (101)&  7214.73  &  1.05    &  41:  &          1.11 &           \\
 &&&&&                                                                         \\
FeII (73) &    7222.39  &  1.1:    &       &          1.11:&           \\
FeII (73) &    7224.47  &  1.2:    &       &          1.22:&     52:   \\
          &           &          &       &               &           \\
 HeI (45) &    7281.35  &  1.70    &  20:  &               &           \\
 &&&&&                                                                        \\
${\rm [CaII] }$ 1F &  7291.46 &   1.60:  &   34: &     2.40:&     68  \\
 &&&&&                                                                        \\
FeII (73) &   7320.70 &   1.15:  &       &           1.22:&     50: \\
${\rm [CaII] }$ 1F & 7323.88 &   1.40:  &   31: &      2.23&      65  \\
 &&&&&                                                                        \\
 MnII (4) &     7330.54 &   1.05:  &       &           1.05:&         \\
 &&&&&                                                                        \\
 MnII (4) &     7353.52 &   1.09:  &       &           1.09:&     50: \\
TiII (101)&     7355.46 &          &       &           1.10:&         \\
 &&&&&                                                               \\
          &     7378?   &   1.08   &       &           1.10&           \\
 &&&&&                                                                         \\
${\rm [FeII] }$ 14F& 7388.16 &   1.11   &   15: &   1.08:&          \\
 &&&&&                                                                         \\
${\rm [VII]}$ 4F  &   7411.9  &   1.06:  &    9: &     1.07&      53:  \\
 &&&&&                                                                         \\
 MnII (4) &     7415.78 &   1.18   &   39: &           1.17&      50:  \\
 &&&&&                                                                         \\
 NI   (3) &     7423.64 &   1.06   &       &           0.82&      80  \\
 &&&&&                                                                        \\
 MnII (4) &     7432.27 &   1.06   &   37  &           1.11&      52  \\
 &&&&&                                                                        \\
 NI   (3) &     7442.29 &   1.07   &   37: &           0.77&      80  \\
 &&&&&                                                                        \\
FeII (73) &     7449.33 &   1.18   &   40: &           1.23&      50  \\
${\rm [FeII] }$ 14F& 7452.5  &   1.07:  &       &           1.10&      68  \\
 &&&&&                                                                        \\
FeII (73) & 7462.39 &   1.32   &   39  &           1.36&      44  \\
          &           &          &       &           0.96:&    105:\\
 &&&&&                                                                         \\
 NI   (3) & 7468.31 &   1.07   &       &           0.66&      80   \\
 &&&&&                                                                         \\
FeII (72) & 7479.69 &   1.10   &   40  &           1.10&      50:  \\
 &&&&&                                                                         \\
 FeII     & 7495.63 &   1.07   &       &           1.02:&          \\
 &&&&&                                                                         \\
FeII (73) & 7515.79 &   1.27   &   37  &           1.39&           \\
\end{supertabular}
\end{center}

\onecolumn
\newpage
\begin{figure}
\includegraphics[angle=-90,width=1.0\textwidth,bb=50 140 450 790,clip]{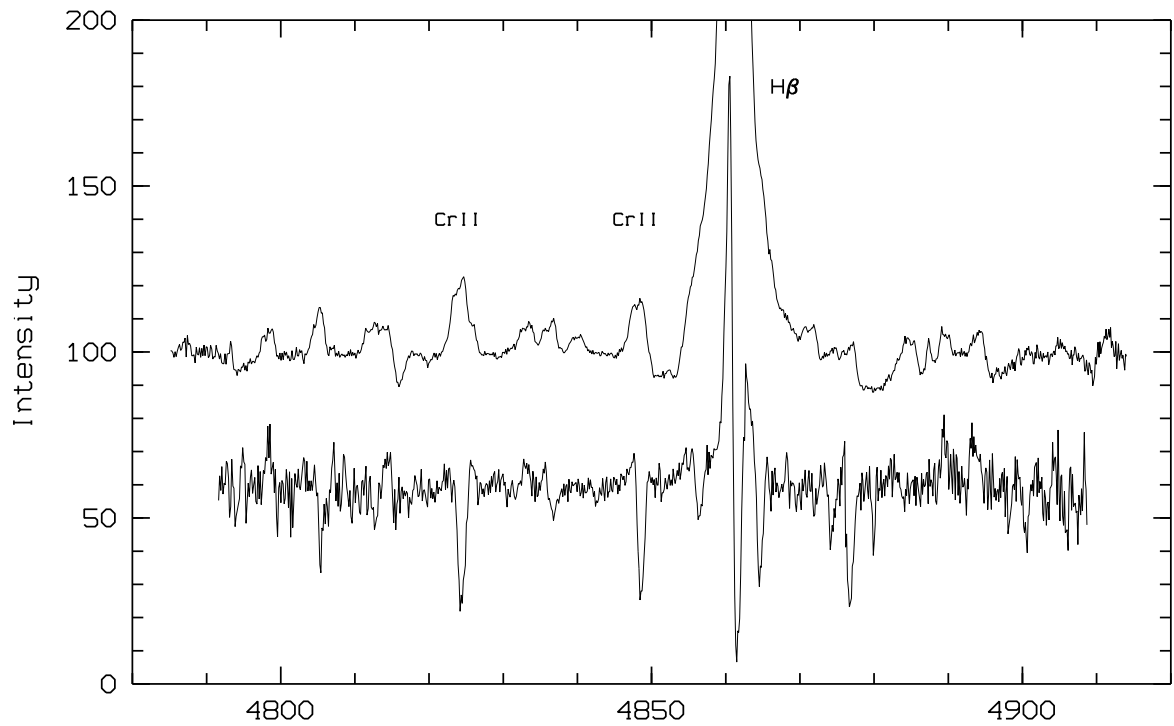}  
\caption{The spectra of MWC\,314 (above) and IRC+10420 (lower). The wavelengths are given in Angstroms, 
        the intensity is normalized to the continuum level}
\end{figure}

\newpage
\setcounter{figure}{2}
\begin{figure}
\includegraphics[angle=-90,width=1.0\textwidth,bb=50 140 450 790,clip]{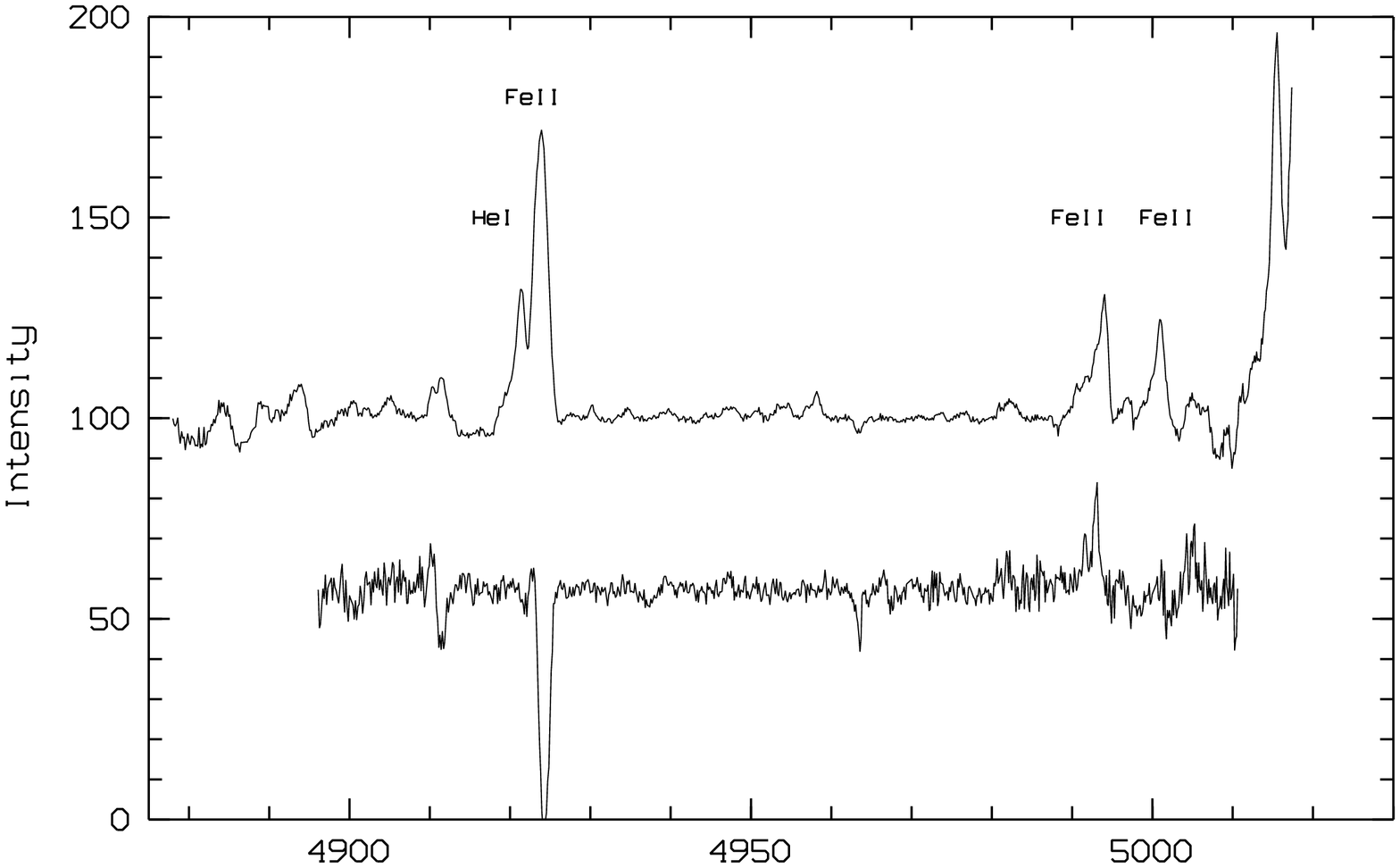}  
\caption{}
\end{figure}

\newpage
\setcounter{figure}{2}
\begin{figure}
\includegraphics[angle=-90,width=1.0\textwidth,bb=50 140 450 790,clip]{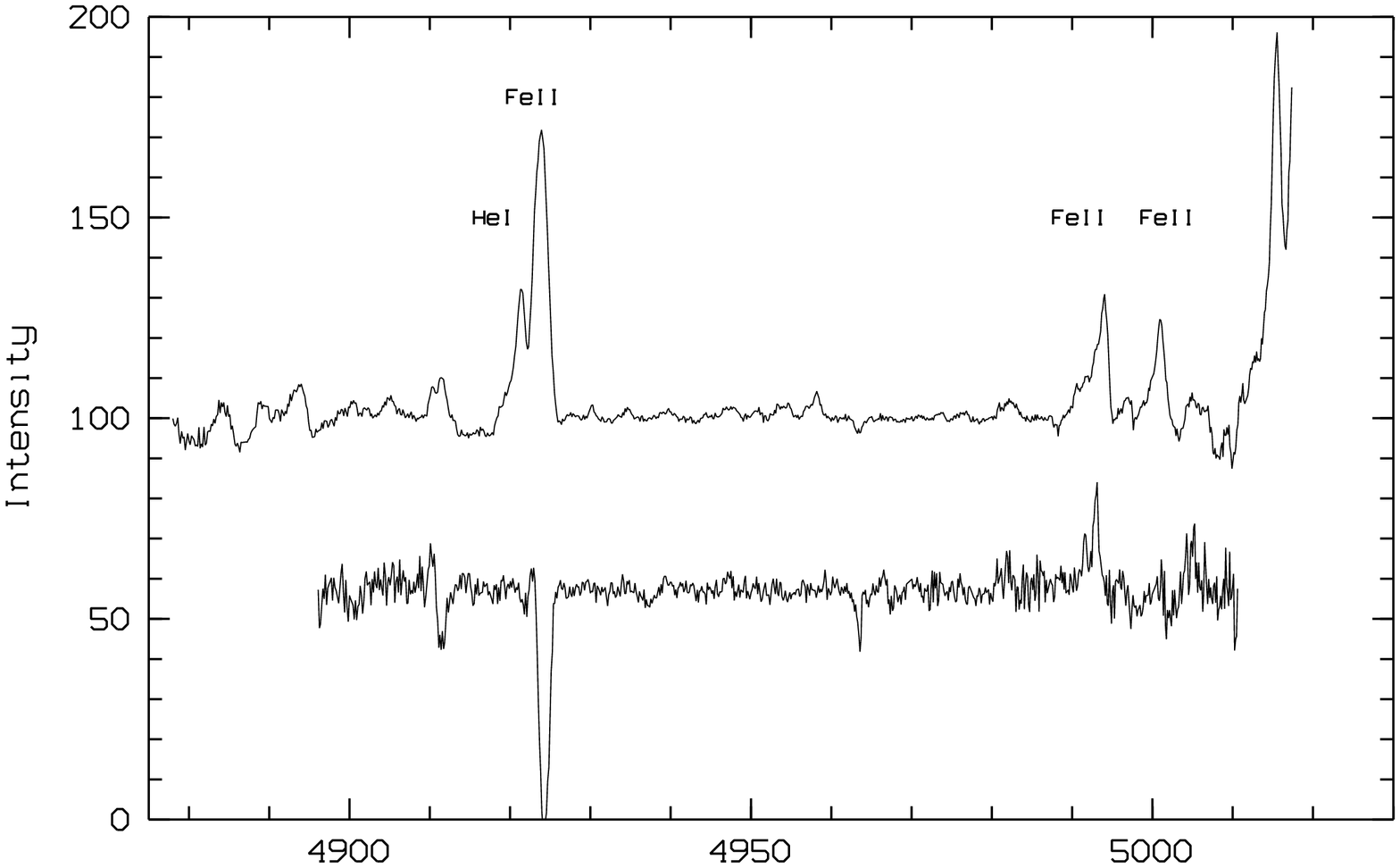}  
\caption{}
\end{figure}

\newpage
\setcounter{figure}{2}
\begin{figure}
\includegraphics[angle=-90,width=1.0\textwidth,bb=50 140 450 790,clip]{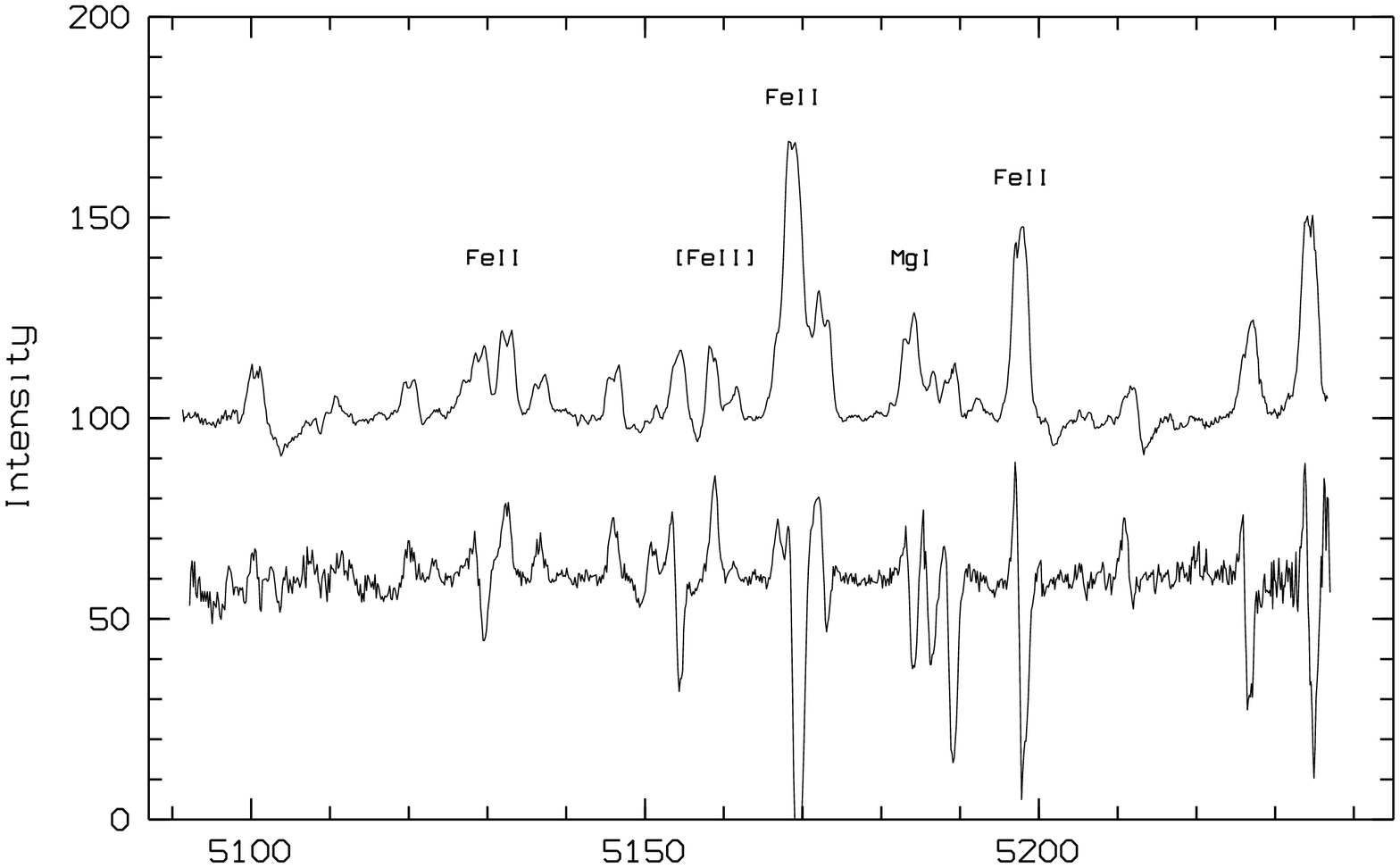}  
\caption{}
\end{figure}

\newpage
\setcounter{figure}{2}
\begin{figure}
\includegraphics[angle=-90,width=1.0\textwidth,bb=50 140 450 790,clip]{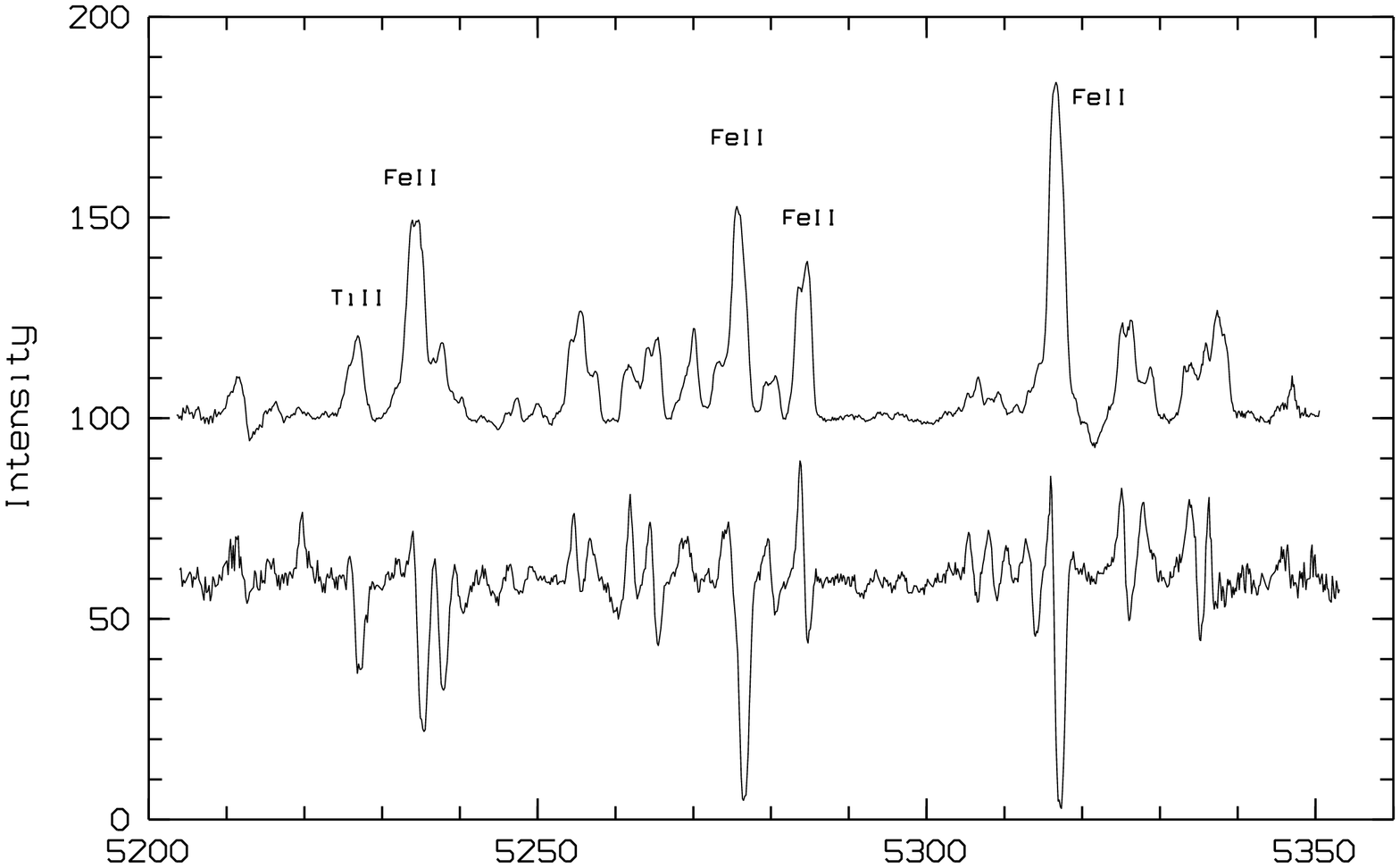}  
\caption{}
\end{figure}

\newpage
\setcounter{figure}{2}
\begin{figure}
\includegraphics[angle=-90,width=1.0\textwidth,bb=50 140 450 790,clip]{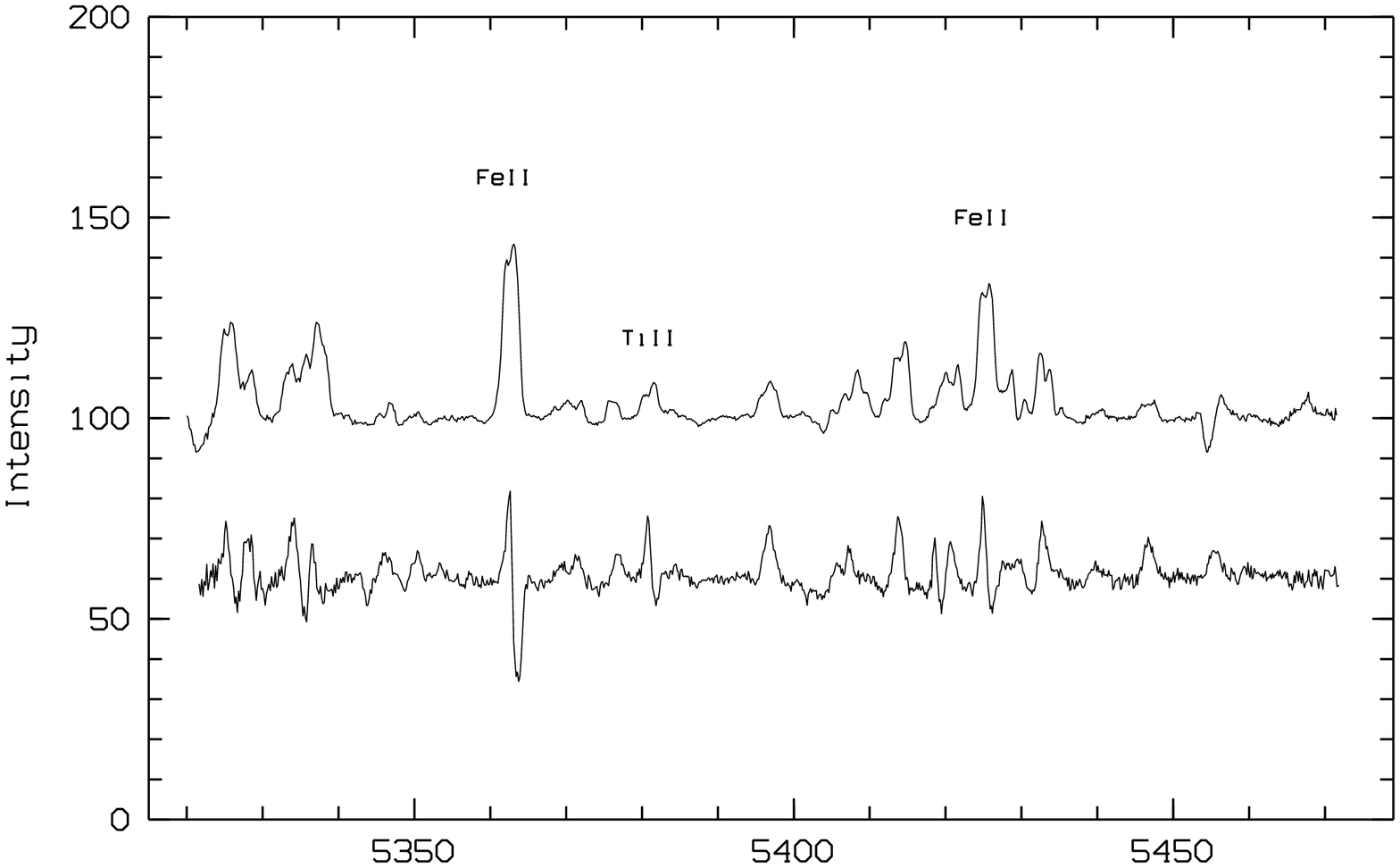}  
\caption{}
\end{figure}

\newpage
\setcounter{figure}{2}
\begin{figure}
\includegraphics[angle=-90,width=1.0\textwidth,bb=50 140 450 790,clip]{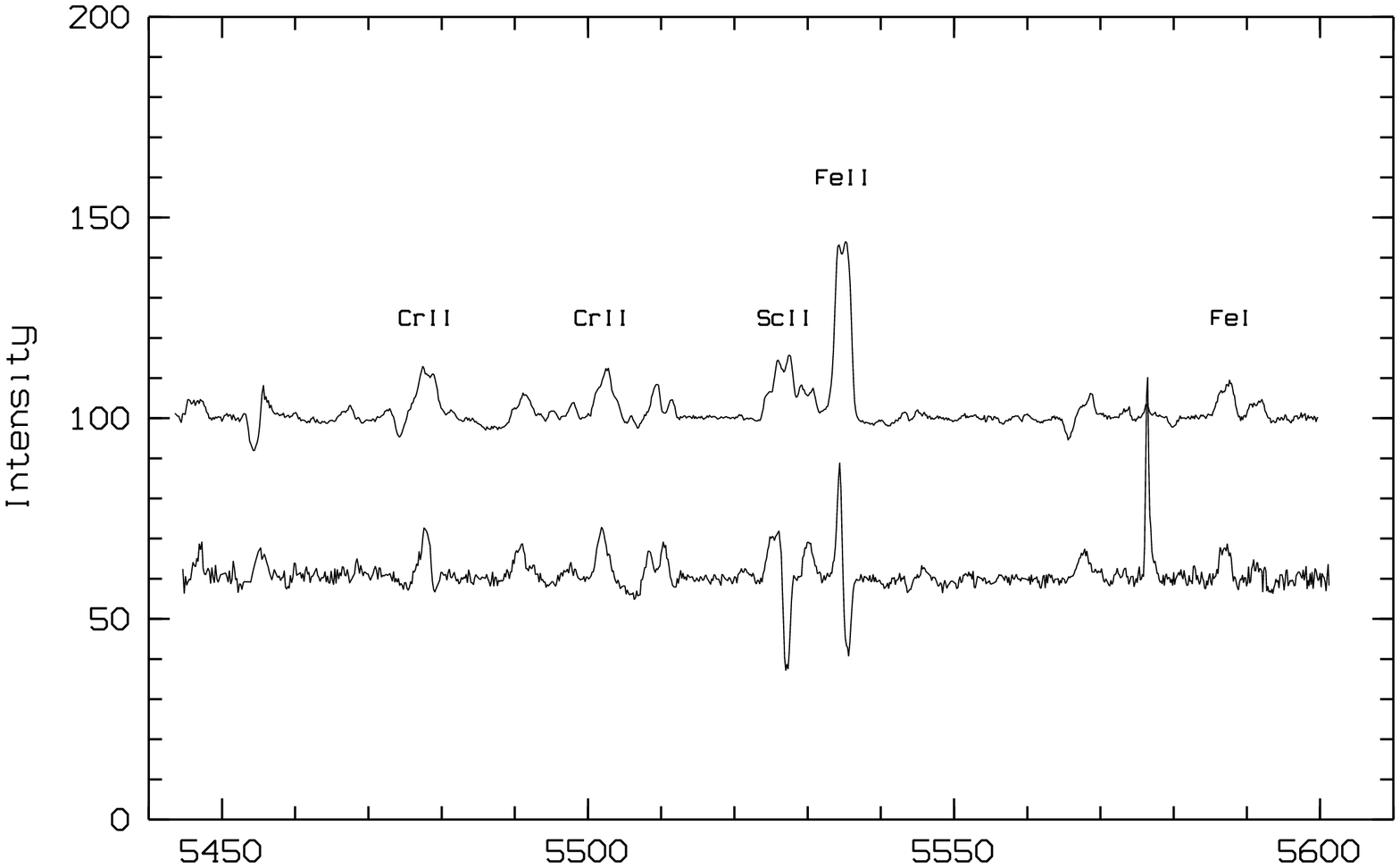}  
\caption{}
\end{figure}

\newpage
\setcounter{figure}{2}
\begin{figure}
\includegraphics[angle=-90,width=1.0\textwidth,bb=50 140 450 790,clip]{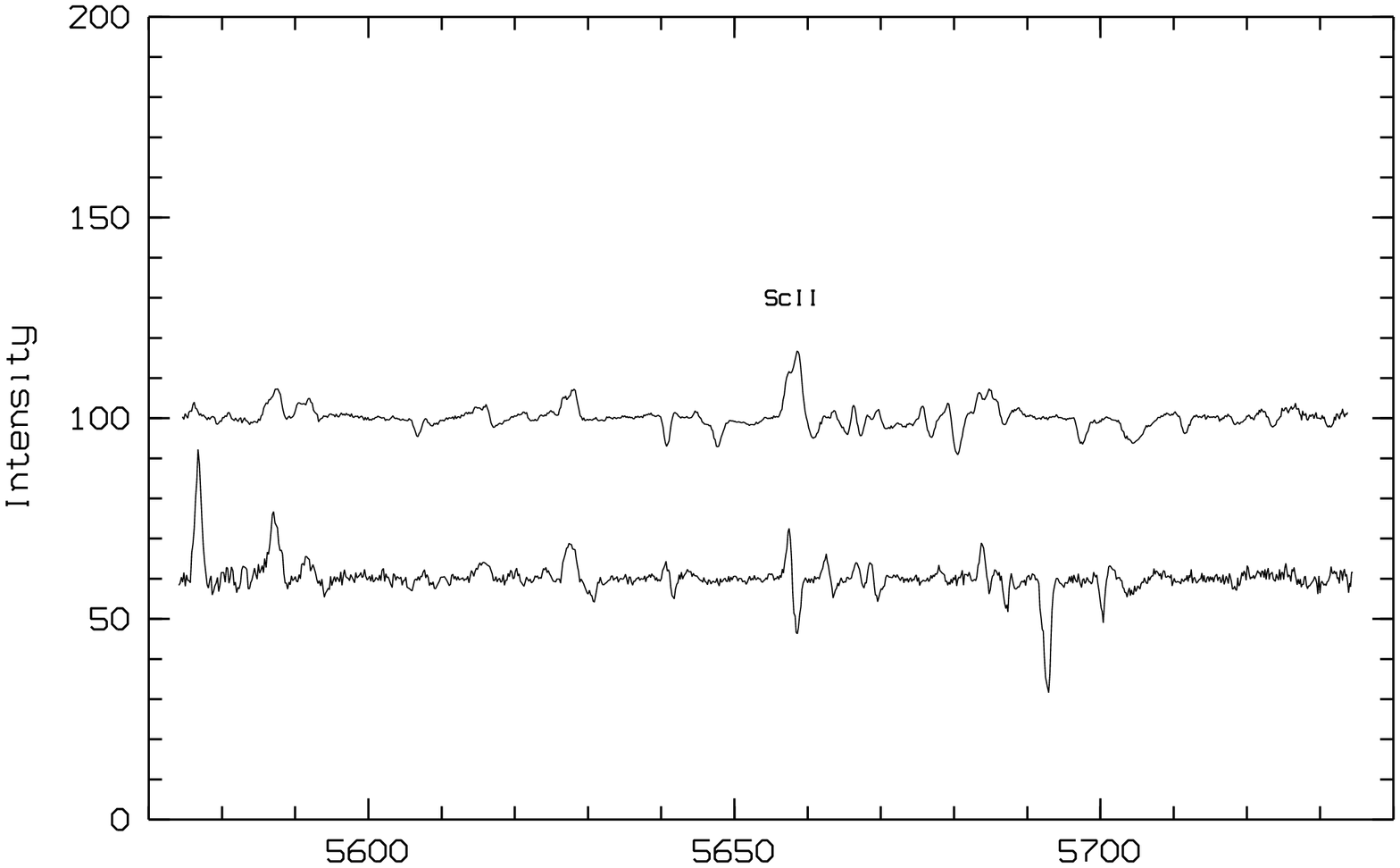}  
\caption{}
\end{figure}

\newpage
\setcounter{figure}{2}
\begin{figure}
\includegraphics[angle=-90,width=1.0\textwidth,bb=50 140 450 790,clip]{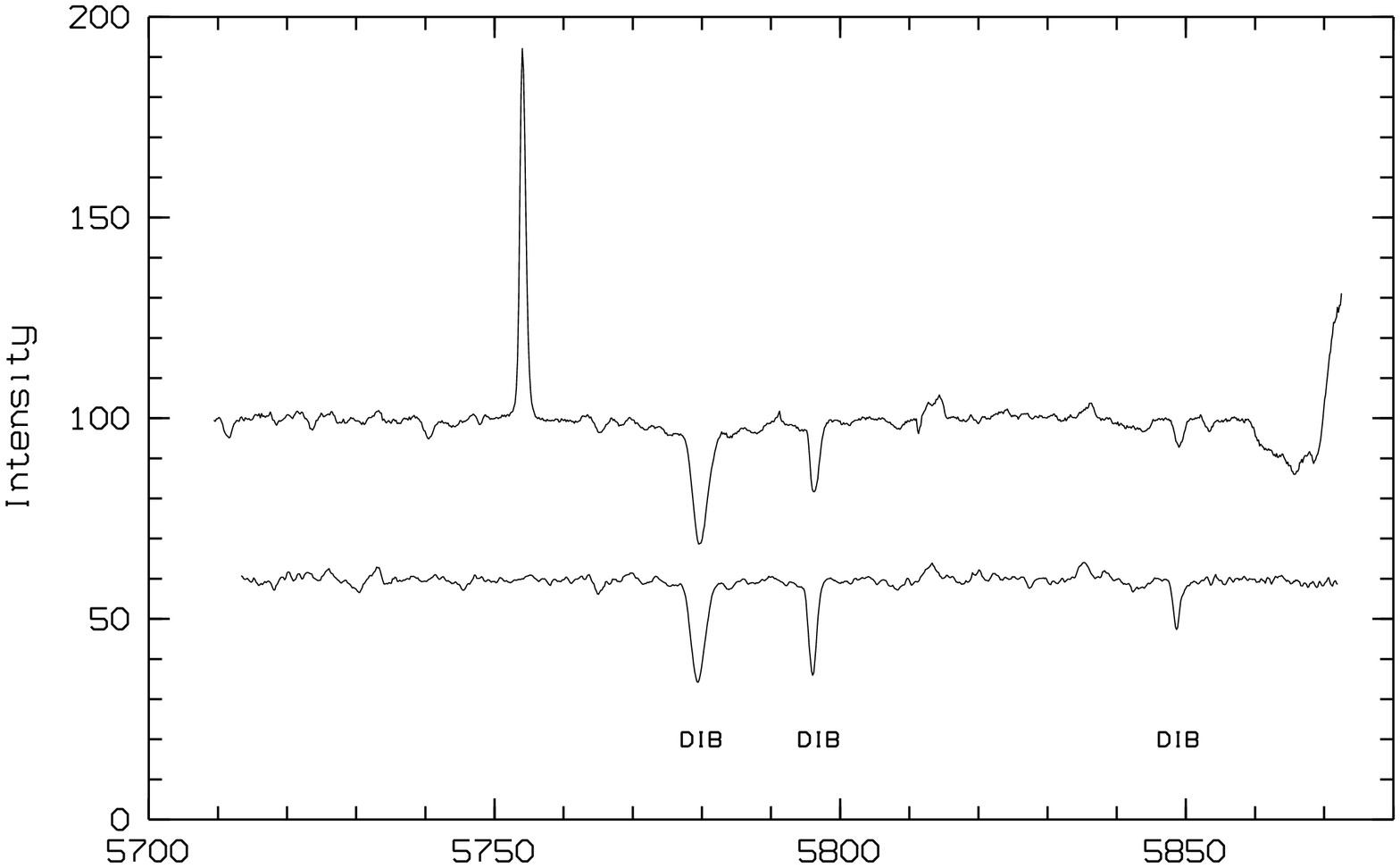}  
\caption{}
\end{figure}

\newpage
\setcounter{figure}{2}
\begin{figure}
\includegraphics[angle=-90,width=1.0\textwidth,bb=50 140 450 790,clip]{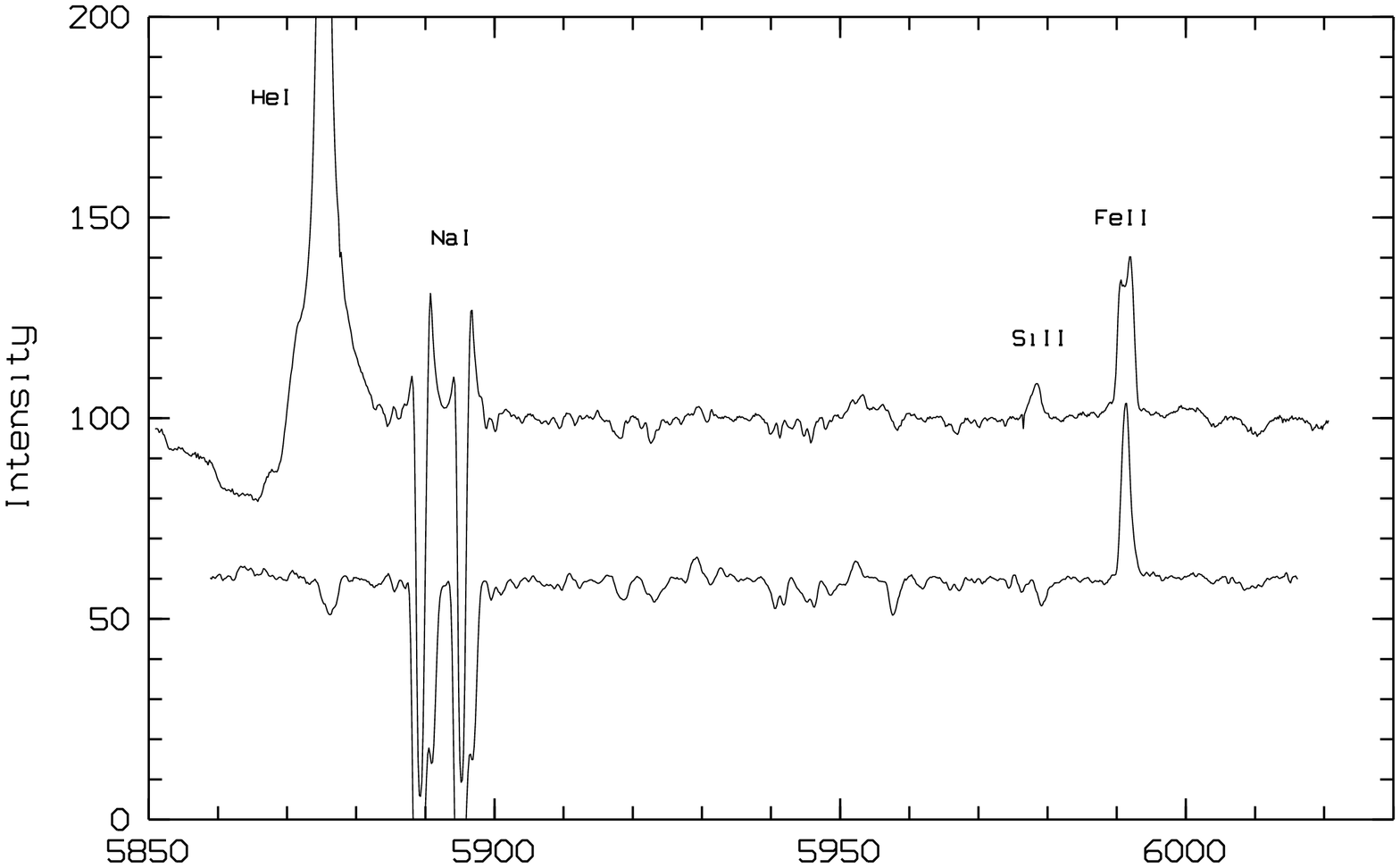}  
\caption{}
\end{figure}

\newpage
\setcounter{figure}{2}
\begin{figure}
\includegraphics[angle=-90,width=1.0\textwidth,bb=50 140 450 790,clip]{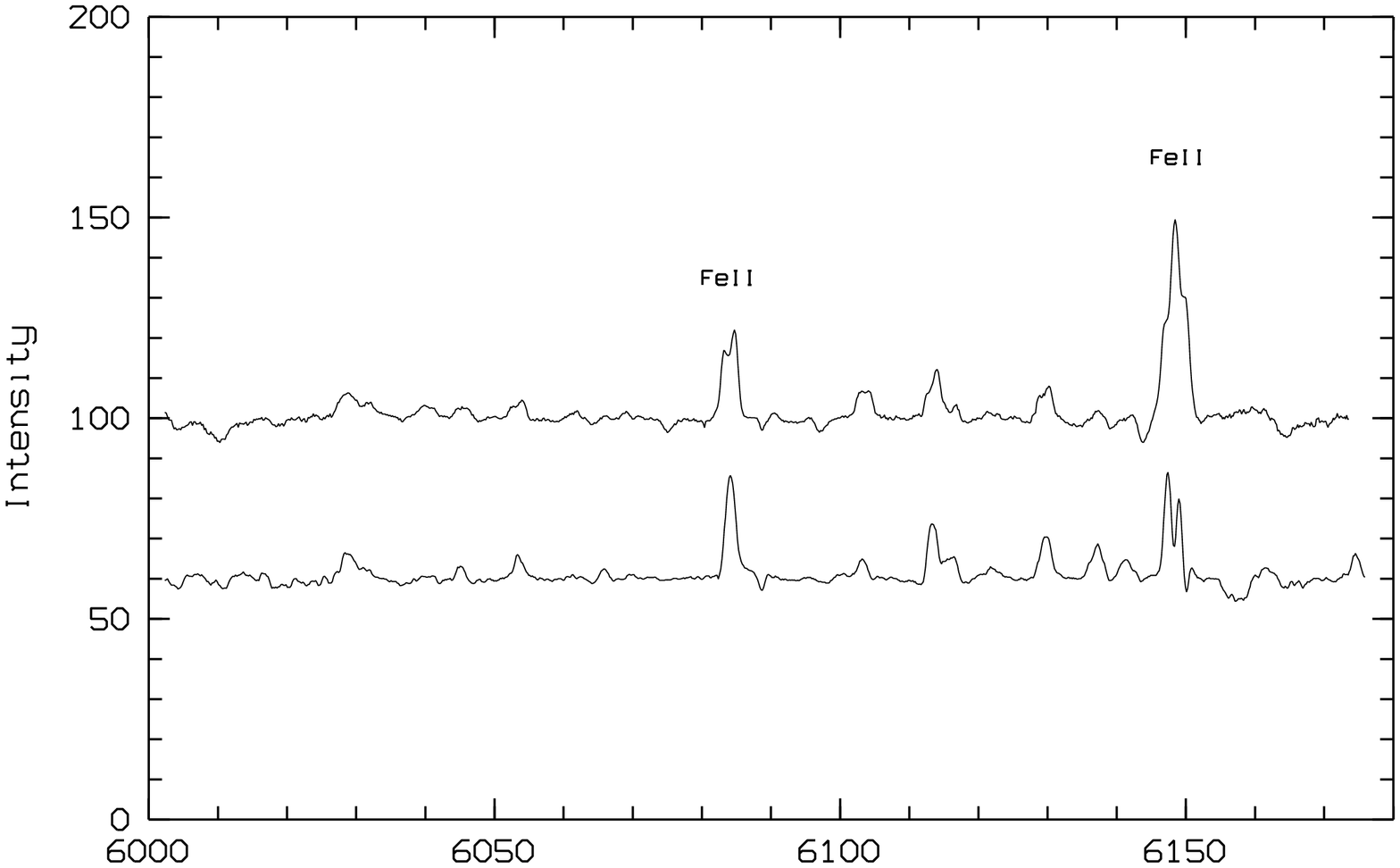}  
\caption{}
\end{figure}

\newpage
\setcounter{figure}{2}
\begin{figure}
\includegraphics[angle=-90,width=1.0\textwidth,bb=50 140 450 790,clip]{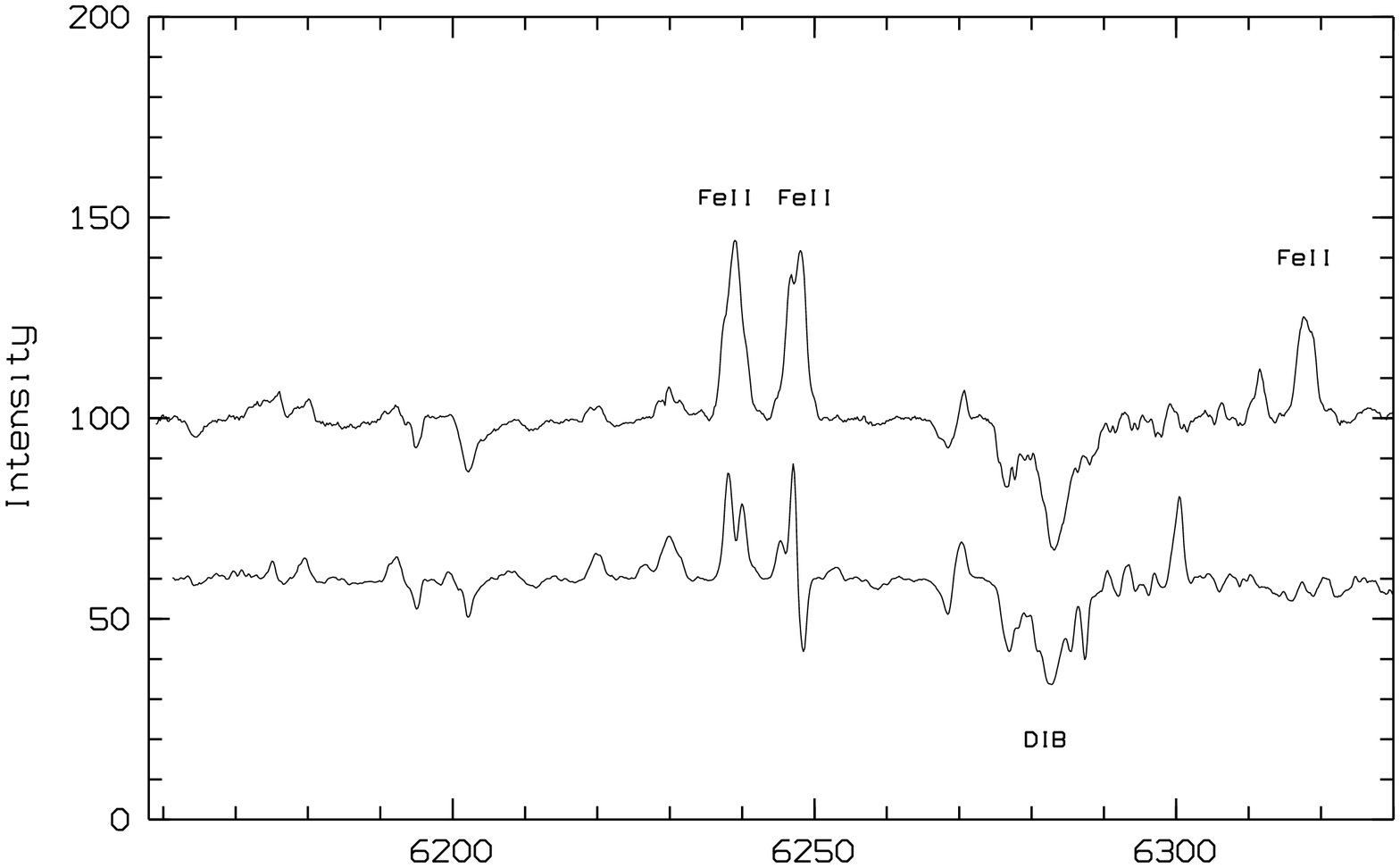}  
\caption{}
\end{figure}

\newpage
\setcounter{figure}{2}
\begin{figure}
\includegraphics[angle=-90,width=1.0\textwidth,bb=50 140 450 790,clip]{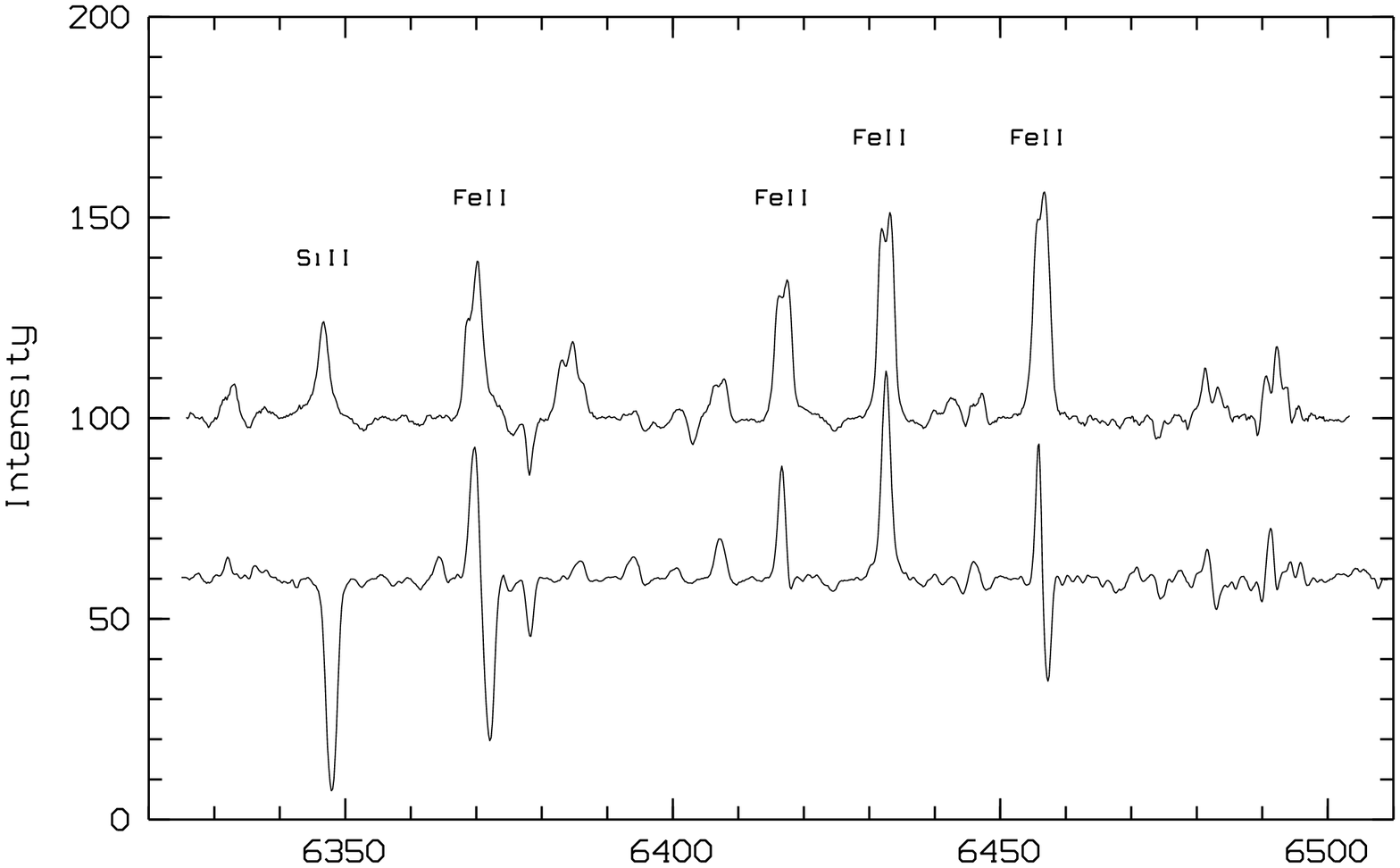}  
\caption{}
\end{figure}

\newpage
\setcounter{figure}{2}
\begin{figure}
\includegraphics[angle=-90,width=1.0\textwidth,bb=50 140 450 790,clip]{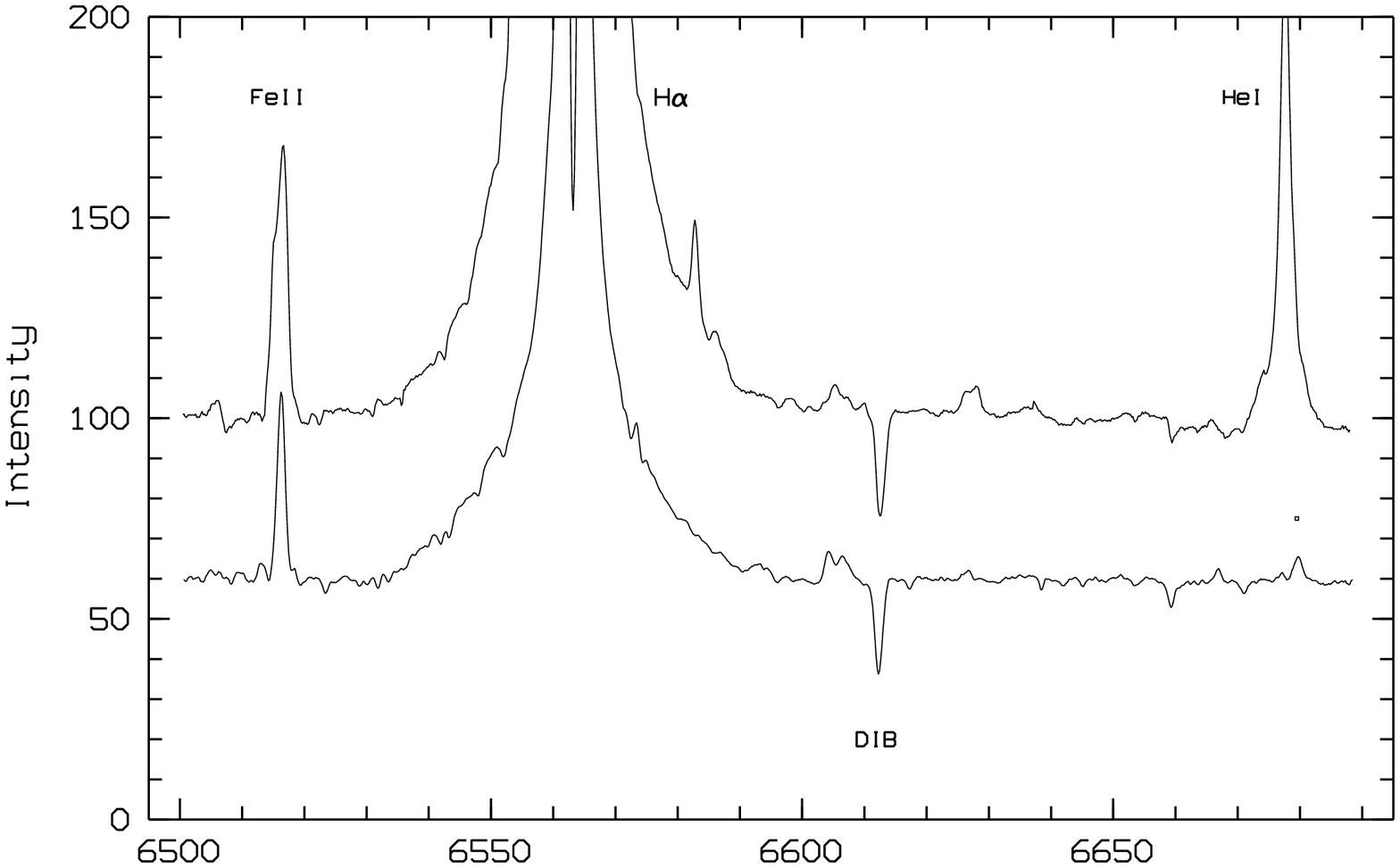}  
\caption{}
\end{figure}

\newpage
\setcounter{figure}{2}
\begin{figure}
\includegraphics[angle=-90,width=1.0\textwidth,bb=50 140 450 790,clip]{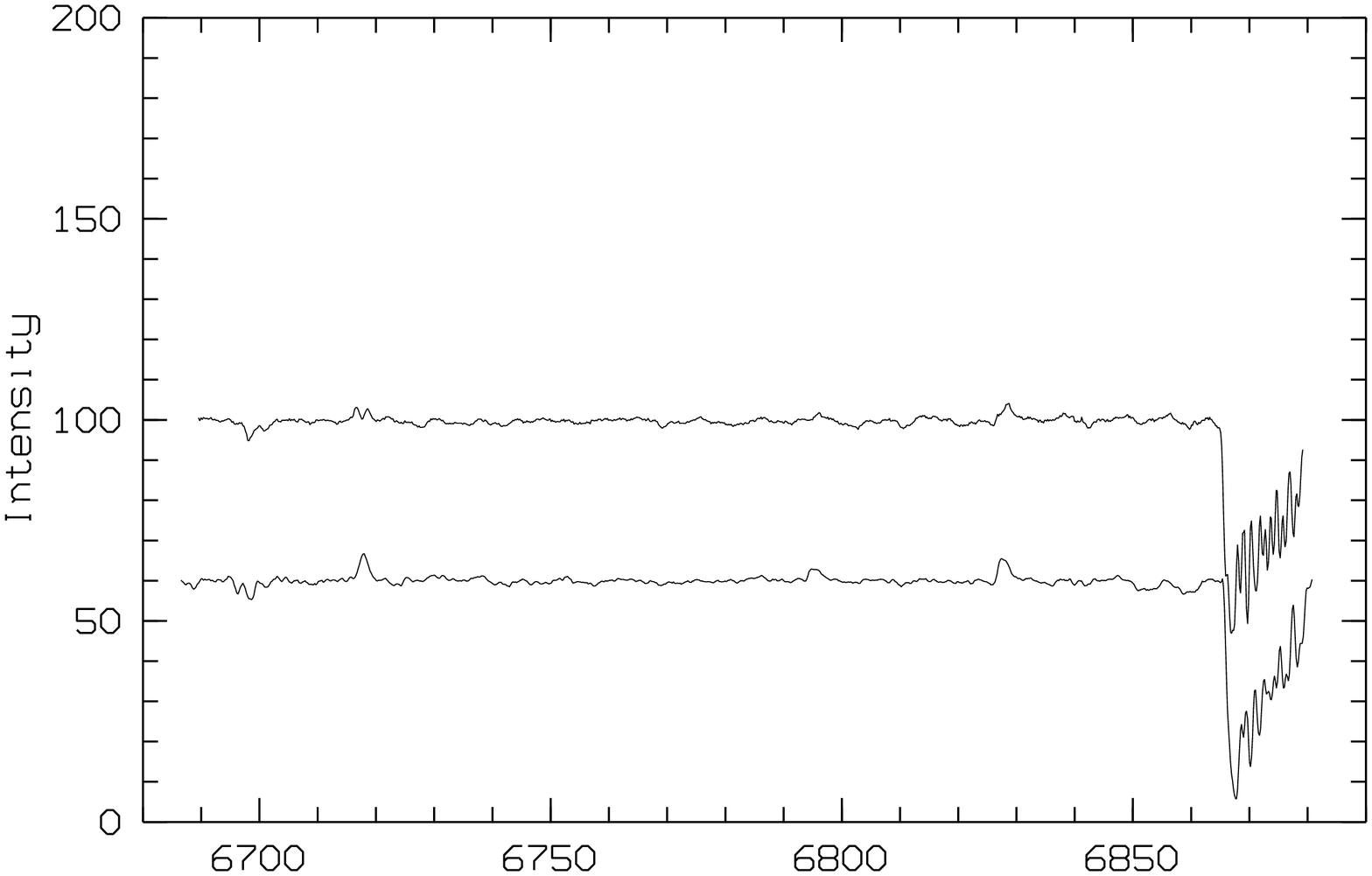}  
\caption{}
\end{figure}

\newpage
\setcounter{figure}{2}
\begin{figure}
\includegraphics[angle=-90,width=1.0\textwidth,bb=50 140 450 790,clip]{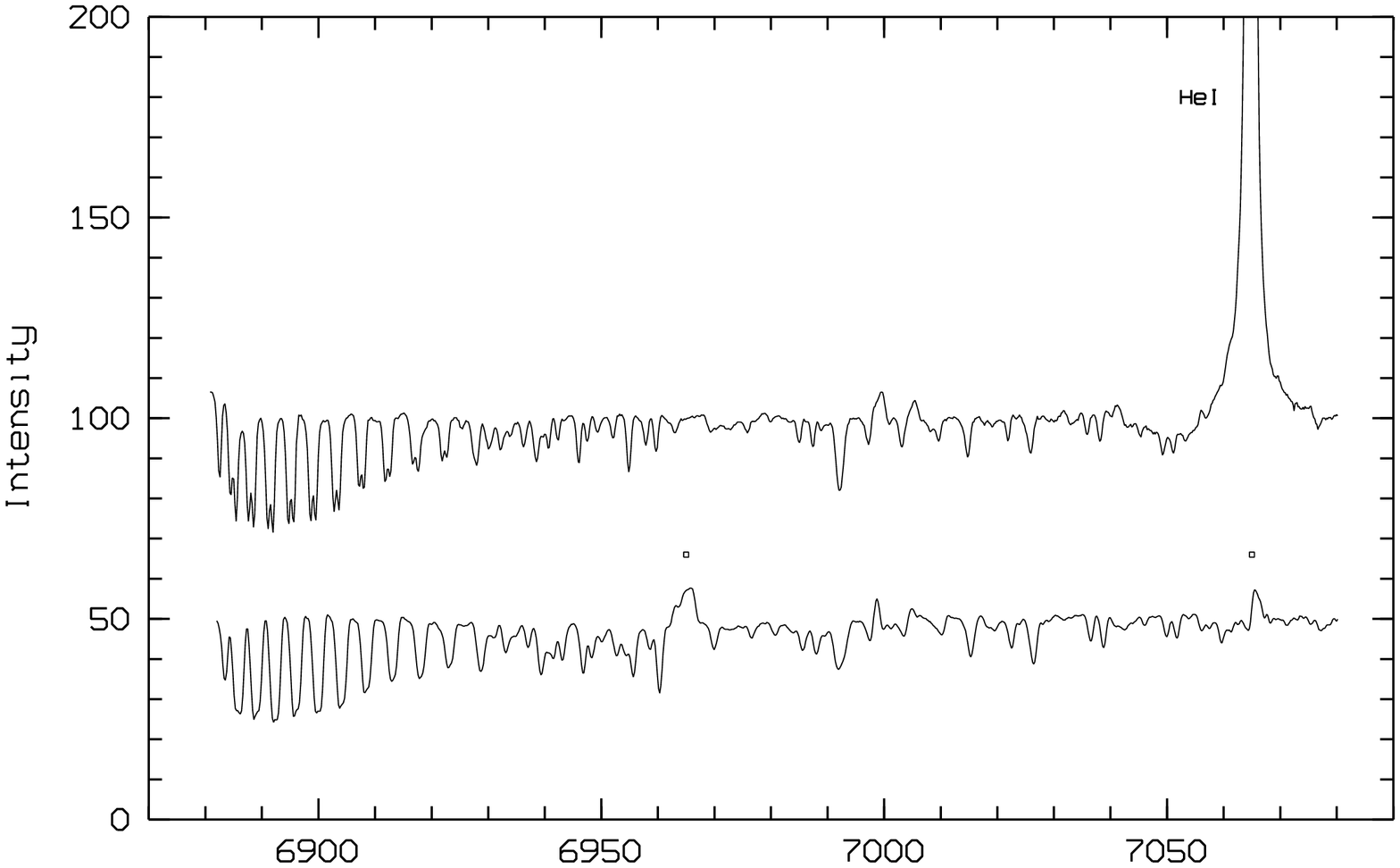}  
\caption{}
\end{figure}

\newpage
\setcounter{figure}{2}
\begin{figure}
\includegraphics[angle=-90,width=1.0\textwidth,bb=50 140 450 790,clip]{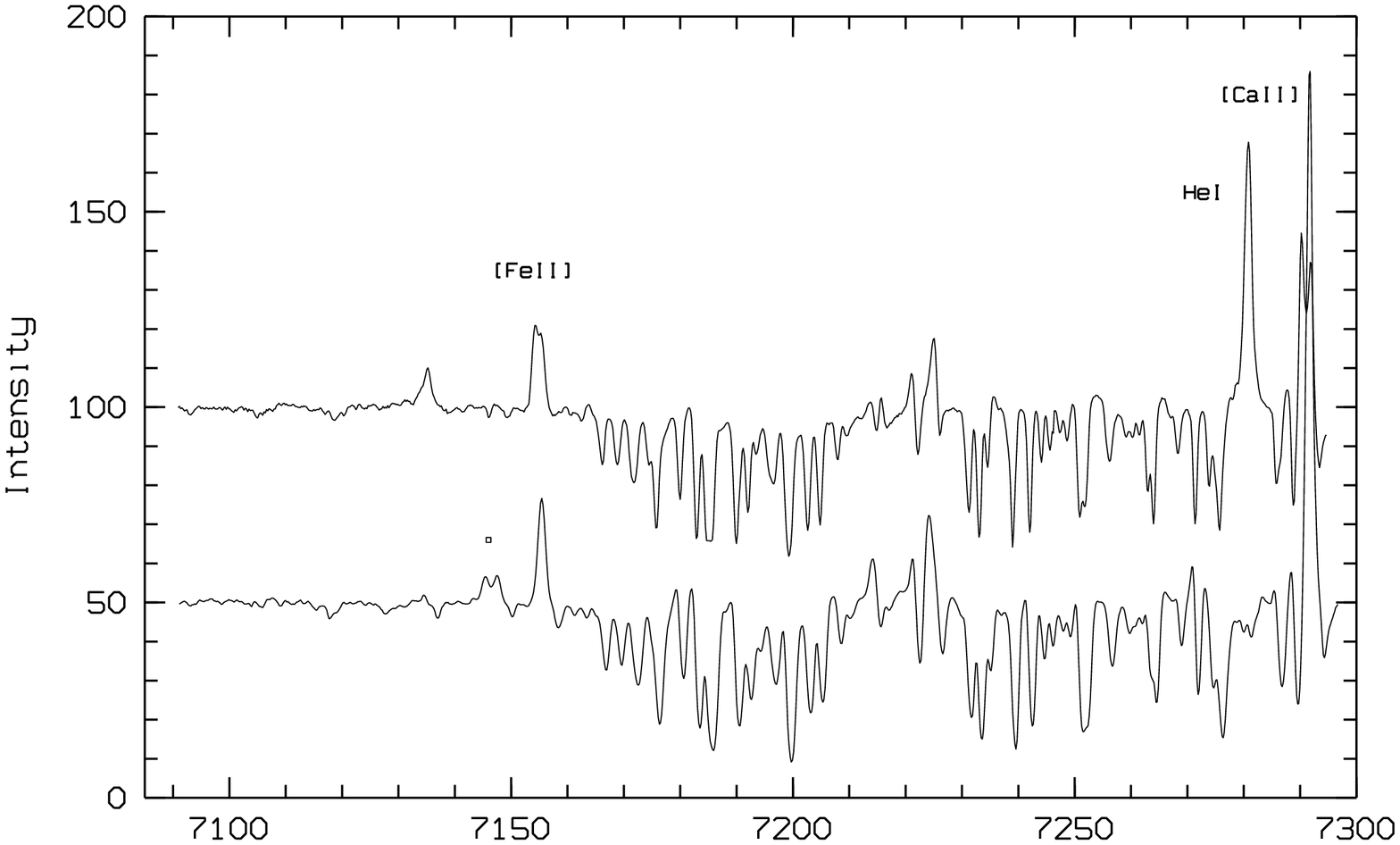}  
\caption{}
\end{figure}

\newpage
\setcounter{figure}{2}
\begin{figure}
\includegraphics[angle=-90,width=1.0\textwidth,bb=50 140 450 790,clip]{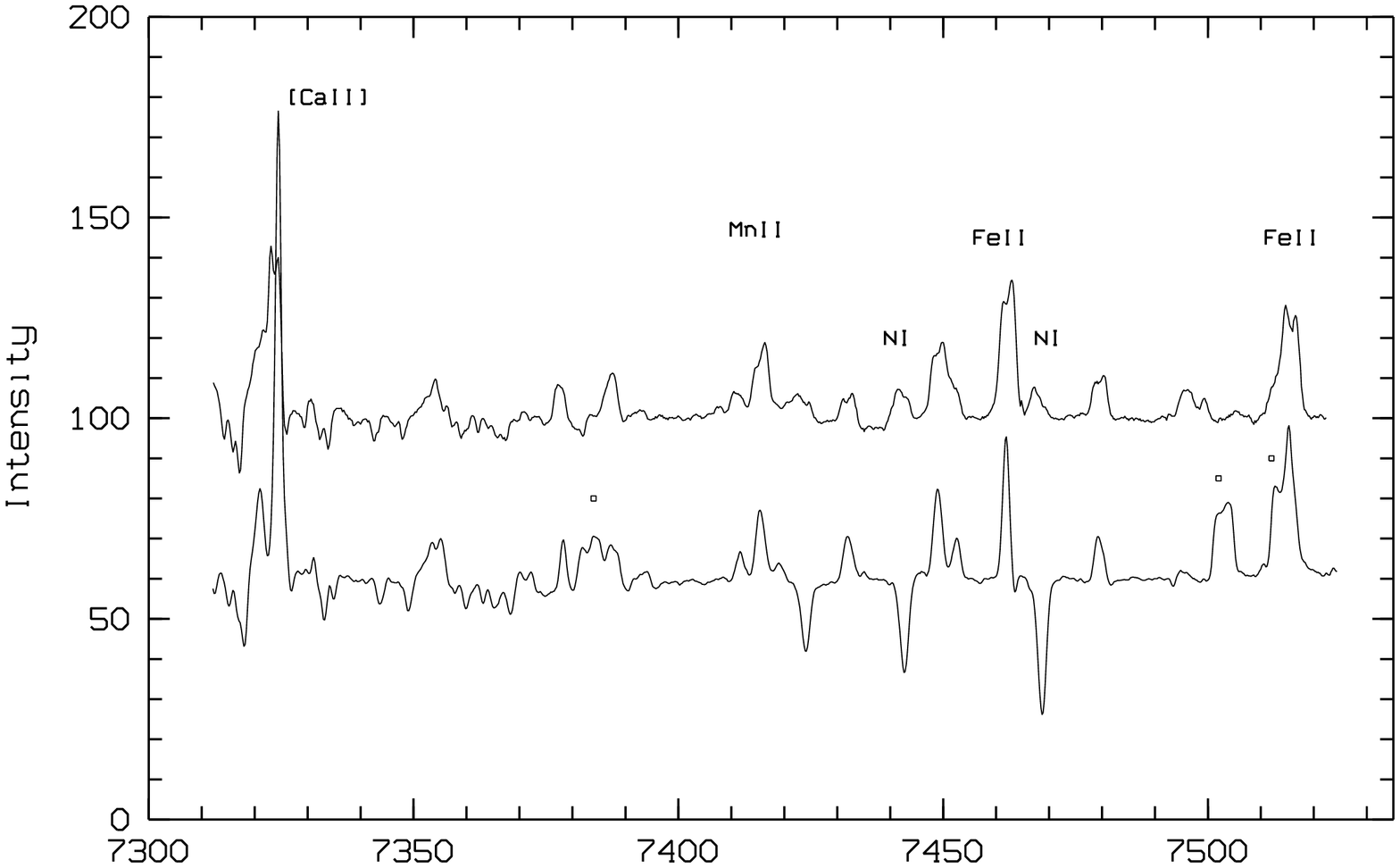}  
\caption{}
\end{figure}


\begin{thebibliography}{}
\bibitem{}
Avedisova V.A., 1996, \alet\ {\bf 22}, 497
\bibitem{}
Klochkova V.G., Chentsov E.L., Panchuk V.E., 1997, \mnras\ {\bf 292}, 19
\bibitem{}
Lamers H.J.G.L.M., Zickgraph F-J., Winter D. de, Houziaux L., Zorec J.,
1998, \aaa\ {\bf 340}, 117
\bibitem{}
Miroshnichenko A.S., Fremat Y., Houziaux L., Andrillat Y., Chentsov E.L.,
Klochkova V.G., 1998, \aas\ {\bf 131}, 469
\bibitem{}
Oudmaiyer R.D., 1995, PhD thesis, Univ. Groningen
\bibitem{}
Oudmaiyer R.D., 1998, \aas\ {\bf 129}, 541
\bibitem{}
Panchuk V.E., Najdenov I.D., Klochkova V.G., Ivanchuk A.B., Yermakov S.V.,
Murzin V.A., 1998, \bsao\ {\bf 44}, 127
\end{thebibliography}
\end{document}